\documentstyle[12pt]{article}
\advance \topmargin by -\headheight
\advance \topmargin by -\headsep

\evensidemargin \oddsidemargin
\begin{document}

\topmargin -.6in

%
%
\def\rf#1{(\ref{eq:#1})}
\def\lab#1{\label{eq:#1}}
\def\nonu{\nonumber}
\def\br{\begin{eqnarray}}
\def\er{\end{eqnarray}}
\def\be{\begin{equation}}
\def\ee{\end{equation}}
\def\foot#1{\footnotemark\footnotetext{#1}}
\def\lb{\lbrack}
\def\rb{\rbrack}
\def\llangle{\left\langle}
\def\rrangle{\right\rangle}
\def\blangle{\Bigl\langle}
\def\brangle{\Bigr\rangle}
\def\llbrack{\left\lbrack}
\def\rrbrack{\right\rbrack}
\def\lcurl{\left\{}
\def\rcurl{\right\}}
\def\({\left(}
\def\){\right)}
\def\v{\vert}
\def\bv{\bigm\vert}
\def\Bgv{\;\Bigg\vert}
\def\bgv{\bigg\vert}
\def\lskip{\vskip\baselineskip\vskip-\parskip\noindent}
\relax

\def\tr{\mathop{\rm tr}}
\def\Tr{\mathop{\rm Tr}}
\def\partder#1#2{{{\partial #1}\over{\partial #2}}}
\def\funcder#1#2{{{\delta #1}\over{\delta #2}}}
\def\me#1#2{\left\langle #1\right|\left. #2 \right\rangle}
\def\a{\alpha}
\def\b{\beta}
\def\d{\delta}
\def\D{\Delta}
\def\eps{\epsilon}
\def\vareps{\varepsilon}
\def\g{\gamma}
\def\G{\Gamma}
\def\grad{\nabla}
\def\h{{1\over 2}}
\def\l{\lambda}
\def\L{\Lambda}
\def\m{\mu}
\def\n{\nu}
\def\o{\over}
\def\om{\omega}
\def\O{\Omega}
\def\p{\phi}
\def\P{\Phi}
\def\vp{\varphi}
\def\va{\vartheta}
\def\pa{\partial}
\def\pr{\prime}
\def\qb{{\bar q}}
\def\qt{{\tilde q}}
\def\ra{\rightarrow}
\def\s{\sigma}
\def\S{\Sigma}
\def\t{\tau}
\def\th{\theta}
\def\Th{\Theta}
\def\ti{\tilde}
\def\wti{\widetilde}
\def\veps{\varepsilon}
\def\tg{\bigtriangleup}
%
\def\lie{{\cal G}}
\def\dlie{{\cal G}^{\ast}}
\def\elie{{\widetilde \lie}}
\def\edlie{{\elie}^{\ast}}
\def\hatlie{{\hat {\lie}}}
\def\hlie{{\cal H}}
\def\wlie{{\widetilde \lie}}
\def\sur{$SU(r+1)$}
\def\spr{$Sp(r)$}
\def\soe{$SO(2r)$}
\def\sod{$SO(2r+1)$}
\def\gtwo{$G_2$}
\def\ffour{$F_4$}
\def\esix{$E_6$}
\def\eeight{$E_8$}
\def\sw{w_{\infty}}
\def\bw{W_{\infty}}
\def\osw{w_{1+ \infty}}
\def\obw{W_{1+\infty}}

\def\rlx{\relax\leavevmode}
\def\inbar{\vrule height1.5ex width.4pt depth0pt}
\def\IZ{\rlx\hbox{\sf Z\kern-.4em Z}}
\def\IR{\rlx\hbox{\rm I\kern-.18em R}}
%
%
\def\mark{\noindent{\bf Remark.}\quad}
\def\prop{\noindent{\bf Proposition.}\quad}
\def\proof{\noindent{\bf Proof.}\quad}
\newcommand{\nit}{\noindent}
\newcommand{\ct}[1]{\cite{#1}}
\newcommand{\bi}[1]{\bibitem{#1}}
\newtheorem{theor}{Theorem}[section]
%
%
\def\PRL#1#2#3{{\sl Phys. Rev. Lett.} {\bf#1} (#2) #3}
\def\NPB#1#2#3{{\sl Nucl. Phys.} {\bf B#1} (#2) #3}
\def\NPBFS#1#2#3#4{{\sl Nucl. Phys.} {\bf B#2} [FS#1] (#3) #4}
\def\CMP#1#2#3{{\sl Commun. Math. Phys.} {\bf #1} (#2) #3}
\def\PRD#1#2#3{{\sl Phys. Rev.} {\bf D#1} (#2) #3}
\def\PLA#1#2#3{{\sl Phys. Lett.} {\bf #1A} (#2) #3}
\def\PLB#1#2#3{{\sl Phys. Lett.} {\bf #1B} (#2) #3}
\def\JMP#1#2#3{{\sl J. Math. Phys.} {\bf #1} (#2) #3}
\def\PTP#1#2#3{{\sl Prog. Theor. Phys.} {\bf #1} (#2) #3}
\def\SPTP#1#2#3{{\sl Suppl. Prog. Theor. Phys.} {\bf #1} (#2) #3}
\def\AoP#1#2#3{{\sl Ann. of Phys.} {\bf #1} (#2) #3}
\def\PNAS#1#2#3{{\sl Proc. Natl. Acad. Sci. USA} {\bf #1} (#2) #3}
\def\RMP#1#2#3{{\sl Rev. Mod. Phys.} {\bf #1} (#2) #3}
\def\PR#1#2#3{{\sl Phys. Reports} {\bf #1} (#2) #3}
\def\AoM#1#2#3{{\sl Ann. of Math.} {\bf #1} (#2) #3}
\def\UMN#1#2#3{{\sl Usp. Mat. Nauk} {\bf #1} (#2) #3}
\def\FAP#1#2#3{{\sl Funkt. Anal. Prilozheniya} {\bf #1} (#2) #3}
\def\FAaIA#1#2#3{{\sl Functional Analysis and Its Application} {\bf #1} (#2)
#3}
\def\BAMS#1#2#3{{\sl Bull. Am. Math. Soc.} {\bf #1} (#2) #3}
\def\TAMS#1#2#3{{\sl Trans. Am. Math. Soc.} {\bf #1} (#2) #3}
\def\Invm#1#2#3{{\sl Invent. Math.} {\bf #1} (#2) #3}
\def\LMP#1#2#3{{\sl Letters in Math. Phys.} {\bf #1} (#2) #3}
\def\IJMPA#1#2#3{{\sl Int. J. Mod. Phys.} {\bf A#1} (#2) #3}
\def\AdM#1#2#3{{\sl Advances in Math.} {\bf #1} (#2) #3}
\def\RMaP#1#2#3{{\sl Reports on Math. Phys.} {\bf #1} (#2) #3}
\def\IJM#1#2#3{{\sl Ill. J. Math.} {\bf #1} (#2) #3}
\def\APP#1#2#3{{\sl Acta Phys. Polon.} {\bf #1} (#2) #3}
\def\TMP#1#2#3{{\sl Theor. Mat. Phys.} {\bf #1} (#2) #3}
\def\JPA#1#2#3{{\sl J. Physics} {\bf A#1} (#2) #3}
\def\JSM#1#2#3{{\sl J. Soviet Math.} {\bf #1} (#2) #3}
\def\MPLA#1#2#3{{\sl Mod. Phys. Lett.} {\bf A#1} (#2) #3}
\def\JETP#1#2#3{{\sl Sov. Phys. JETP} {\bf #1} (#2) #3}
\def\JETPL#1#2#3{{\sl  Sov. Phys. JETP Lett.} {\bf #1} (#2) #3}
%

\begin{titlepage}
\vspace*{-1cm}
\noindent
November, 1992 \hfill{IFT-P/052/92}\\
\phantom{bla}
\hfill{UICHEP-TH/92-18} \\
\phantom{bla}
\hfill{hep-th/9212086}
\\
\vskip .3in
\begin{center}
{\large\bf Hirota's Solitons in the Affine and the Conformal Affine Toda
Models}
\end{center}

\normalsize
\vskip .4in

\begin{center}
{ H. Aratyn\footnotemark
\footnotetext{Work supported in part by U.S. Department of Energy,
contract DE-FG02-84ER40173 and by NSF, grant no. INT-9015799}}

\par \vskip .1in \noindent
Department of Physics \\
University of Illinois at Chicago\\
801 W. Taylor St.\\
Chicago, Illinois 60607-7059\\
\par \vskip .3in

\end{center}

\begin{center}
{C.P. Constantinidis\footnotemark\footnotetext{Supported by Fapesp},
L.A. Ferreira\footnotemark
\footnotetext{Work supported in part by CNPq}}, J.F. Gomes$^{\,3}$ and
A.H. Zimerman$^{\,3}$

\par \vskip .1in \noindent
Instituto de F\'{\i}sica Te\'{o}rica-UNESP\\
Rua Pamplona 145\\
01405-900 S\~{a}o Paulo, Brazil
\par \vskip .3in

\end{center}

\begin{center}
{\large {\bf ABSTRACT}}\\
\end{center}
\par \vskip .3in \noindent

We use Hirota's  method formulated as a recursive scheme to
construct complete set of soliton solutions for the affine Toda field theory
based on an arbitrary Lie algebra.
Our solutions include a new class of solitons connected with two different
type of degeneracies encountered in the Hirota's perturbation approach.

We also derive an universal mass formula for all Hirota's solutions to the
Affine Toda model valid for all underlying Lie groups.
Embedding of the Affine Toda model in the Conformal Affine Toda model plays a
crucial role in this analysis.

\end{titlepage}

\section{Introduction}
\setcounter{equation}{0}
Toda field theories are among the most important examples to study
integrability and conformal invariance.
They can be classified according to an underlying Lie algebra.
The conformal Toda (CT) theories are conformally invariant and
completely integrable models associated to finite dimensional Lie algebra
$\lie$.
The affine Toda (AT) models can be regarded as the conformal Toda perturbed
in such a way that although conformal invariance is broken, complete
integrability is preserved.
The underlying Lie algebra is the Loop algebra $\hatlie$
and these models are known to possess soliton solutions.
The third class, the
conformal affine Toda (CAT) theories, are obtained by recovering the conformal
invariance in the AT models by introducing two extra fields.  These fields can
be shown to be related to adding central extensions to the Loop algebra leading
to a full Kac-Moody algebra underlying the CAT model.

The interest in the soliton solutions of the Affine Toda (AT) field theoretical
model is partly motivated by the well-known results for the simpler case of the
Sine Gordon field theory and partly by the special status of the AT theory in
view of its interpretation as deformed unitary minimal theory.

This paper is devoted to obtaining,
in a systematic manner, soliton solutions of the AT model
and linking them to solutions and properties of the CAT.  This is accomplished
by using the Hirota's tau-function method which was already successfully been
applied to the AT field theory, although not all solutions were found, to the
cases of the Lie algebras $A_r$ \ct{hollo}, $C_r$  \ct{CFGZ} and other
algebras \ct{mmc}.  The method involves introducing one extra tau-function
apart from those assigned to each AT field.  It was within the context of the
CAT model in \ct{CFGZ} that this fact was made clear in view of the new fields
introduced in order to restore conformal invariance.

There are three main (but related) goals which  we have accomplished in
this paper:
{\em (i)}
A  description of soliton solutions obtained by Hirota's method for
the AT model defined on an arbitrary Lie algebra.
{\em (ii)}
Analysis of the Hirota's method as a perturbation method based on recurrence
relations.
{\em (iii)}
Application of the link between Conformal Affine Toda (CAT) model and its
integrable deformation represented by AT theory to reveal new features of the
latter. One of the results we obtain here from general arguments of conformal
field theory, is a general mass-formula valid for  soliton solutions.
A parallel and complementary approach to find soliton solutions for the AT
model
was developed in \ct{olive3} using the Leznov-Saveliev method, see also
\ct{olive2} for the B\"{a}klund's method approach to the problem in the case of
$A_r$.

The Hirota's method
employed in this paper constitute essentially a perturbative
approach based on recurrence relations which can be conveniently
characterized in terms of the eigenvalue problem of a matrix
$L_{ij} = l^{\psi}_{i}K_{ij} $ where $K_{ij}$ is the extended
Cartan matrix and $l^{\psi}_i$ are integers in the expansion of
the highest coroot ${\psi \o {\psi^2}} $ in terms of simple
coroots ${\a_i \o{\a_i^2}}$ of an algebra $\lie$.  A soliton solution
is assigned to each eigenvalue of $L_{ij}$ and there can be at
most ${\rm rank}\; \lie+ 1$  when no degeneracy is present.  The fact that
the $L_{ij}$ is singular is crucial in truncating the
perturbative series leading to an exact solution.  For the
purpose of this study we have established a simple way
of solving this eigenvalue problem for $L_{ij}$ matrix in terms of Chebyshev
polynomials and their special properties. The situation is
changed however when there is a degeneracy.  There are two
separate sources of degeneracy.  One comes from the degeneracy
of the  matrix $L_{ij}$.  Any
linear combination of the degenerate eigenvectors  gives rise to
a soliton solution within  the Hirota's method.  This feature
clearly demonstrates the nonlinear superposition principle
underlying the method.  Another degeneracy is an intrinsic
feature of Hirota's perturbation scheme itself. The relevant
recurrence relation is based on adding to the $n$-th order
perturbed solution non trivial elements of the kernel of $L_{ij}
- n^2 \lambda \d_{ij}$ (for $\lambda$ an eigenvalue of $L$) which exists in
some cases ($SU(6p)$ and $SP(3p)$ with $p$ integer).  Remarkably the Hirota
series truncates in these cases too, producing a new class of solutions.  It is
generally true that the soliton solutions obtained in the degenerate cases
truncate at higher orders of the tau-functions.

There are two related ways of understanding AT model and its relation to CAT
model. One way of thinking introduces AT model as a deformation of the CAT
model preserving integrability of the theory.  Such a deformation modifies the
underlying $\bw$ symmetry structure of CAT but still maintain enough
symmetry to keep the AT massive structure both attractive and tractable for
physicists.  On the other hand there is also a parallel approach, introduced
first into the literature in \ct{CFGZ}, which views AT model as a  version of
the CAT model with conformal symmetry being gauge-fixed.  This second approach
proves to be very useful in this paper defining the precise limit in which one
model goes into another.  One particular consequence of the connection between
the CAT and AT models is a special simple relation between their respective
Energy-Momentum (EM) tensors.  The modified EM tensor of CAT model \ct{AFGZ} in
some special limit consists of a sum of EM tensor of AT model and the extra
surface term.  Since there is no mass scale in the CAT model the only
contribution to the soliton masses of the AT model comes from the pure surface
term which ensures enormous simplicity of the final universal expression for
the soliton masses.

The outline of the paper is as follows.
In Section $2$ we discuss a connection between AT and CAT models and relation
between their EM tensors.
In Section $3$ the Hirota's method is discussed as a perturbation approach and
the general features of this perturbation and recurrence relations
are described.  We also derive the general mass formula and
discuss topological charges.
In Sections $4$, $5$, $6$ and $7$ this theory is used to find and describe
soliton solutions of the AT model with the underlying Lie algebras
$A_{r}= SU(r+1)$, $C_{r}=Sp(r)$, $D_r = SO(2r)$ and $B_r = SO(2r+1)$,
respectively.
In Sections $8$, $9$ and $10$ this analysis is repeated for AT models with
exceptional Lie algebras $G_2$, $F_4$ and finally we make a few comments for
the $E_n$ with $n=6,7,8$.  In Appendix A we list for reader's convenience a
number of technical properties of Chebyshev's polynomials.

\section{The CAT and AT Models}
\setcounter{equation}{0}
The equations of motion of the CAT model associated with a simple Lie algebra
$\lie$ are given by \ct{CFGZ,AFGZ,BB}:
\br
\pa_{-} \pa_{+} \vp^a &=& \, {1 \o \qb}\( q^{a} e^{\qb K_{ab} \vp^b}
- \,q^{0} l^{\psi}_{a}  e^{- \qb K_{\psi b} \vp^b } \) e^{\qb \eta}
\lab{todaone} \\
\pa_{-} \pa_{+} \eta &=& 0 \lab{todatwo} \\
\pa_{-} \pa_{+} \nu &=&   {2 \o \psi^2}
\, {q^{0}\o \qb} e^{-\qb K_{\psi b} \vp^{b}} e^{\qb \eta}
\lab{todathree}
\er
where $K_{ab}=2 \a_a.\a_b/{\a_b^2}$ is the Cartan Matrix of $\lie$,
$a,b=1,...$, ${\rm rank} \,\lie\equiv r$, $\psi$ is the highest root of $\lie$,
$K_{\psi b}=2 \psi .\a_b/{\a_b^2}$, $l^{\psi}_{a}$ are positive integers
appearing in the expansion ${\psi \over
\psi^{2}} = l^{\psi}_{a} {\a_{a} \over \a^{2}_{a}}$, where $\a_a$ are the
simple roots of $\lie$ and $q^a$, $q^0$ and $\qb$ are coupling constants.
The derivatives $\pa_{\pm}$ are w.r.t. the light cone coordinates
$x^{\pm} = x \pm t$.

These equations are invariant under the conformal transformations
\br
x_{+} \rightarrow \tilde{x}_{+} = f(x_{+})
 \, \, \, , \, \, \, \,
x_{-} \rightarrow \tilde{x}_{-} = g(x_{-}) \lab{ge}
\er
and
\br
e^{-\vp^a (x_+,x_-)} &\to&
e^{-\tilde{\vp}^a(\tilde{x}_+,\tilde{x}_-)} = e^{-\vp^a (x_+,x_-)}
\lab{fi} \\
e^{-\nu (x_+,x_-)} &\to&
e^{-\tilde{\nu}(\tilde{x}_+,\tilde{x}_-)}= ({df \over dx_+})^{B}
({dg \over dx_-})^{B} e^{-\nu (x_+,x_-)} \lab{ni}\\
e^{-\eta (x_+,x_-)} &\to&
e^{-\tilde{\eta}(\tilde{x}_+,\tilde{x}_-)} =({df \over dx_+})^{{1\over \qb}}
({dg \over dx_-})^{{1 \o \qb}} e^{-\eta (x_+,x_-)}   \lab{mi}
\er
where $f$ and $g$ are analytic functions and $B$ is an arbitrary number.
Therefore $e^{\varphi^a}$ are scalars under conformal transformations
and $e^{ \nu} $ can also be arranged to be a scalar by setting $B=0$ \ct{CFGZ}.

The equations \rf{todaone}-\rf{todathree} can be written in the form of a
zero curvature condition (for the associated linear system)~\ct{CFGZ,BB,AFGZ}
\be
\pa_{+} A_{-} - \pa_{-} A_{+} + \lb A_{+} , A_{-} \rb = 0
\ee
where
\br
A_{+} = \pa_{+} \Phi + e^{ad \Phi} {\cal E}_{+} \, \, \, , \, \, \,
A_{-} = - \pa_{-} \Phi + e^{-ad \Phi} {\cal E}_{-}
\er
and
\br
\Phi &=& {\qb \over 2}\left( \sum_{a=1}^{rank \lie} \varphi^a H_a^0 + \eta T_3
+ \nu C \right)
\lab{Phi}\\
{\cal E}_{+} &=& \sum_{a=1}^{rank \lie}  E_{\a_a}^0 +
 E_{-\psi}^1 \, \, \, , \, \, \, {\cal E}_{-} = \sum_{a=1}^{rank
\lie} q^a E_{-\a_a}^0 +
q^0 E_{\psi}^{-1}
\er
where $T_3 = 2 {\hat \delta}. H^0 + h D$, with ${\hat \delta}={1\o 2}
\sum_{\a > 0}{\a \over {\a^2}}$ (with $\a$ being the positive roots of $\lie$),
is the generator used to perform the so called homogeneous grading of a
Kac-Moody algebra. $H^0_a$, $D$ and $C$ are the generators of the Cartan
subalgebra of the affine Kac-Moody algebra $\hatlie$ associated to $\lie$.
$E_{a_a}^0$ ($E_{-\a_a}^0$), $a=1,2,\ldots$, ${\rm rank} \,\lie=r$,
$E_{-\psi}^{1}$ ($E_{\psi}^{-1}$) are the positive (negative) simple root step
operators of $\hatlie$.

The Lagrangian for the CAT model is given by
\br
{\cal L} = {\qb^2 \o 4} \sum_{a,b=1}^r {2 \o \a^2_a} K_{ab}
\pa_{\rho} \vp^a \pa^{\rho} \vp^b +
{\qb^2 \o 2} \sum_{a=1}^r {2 \o \a^2_a} \pa_{\rho } \vp^a \pa^{\rho} \eta
+ \qb^2 {h \o 2} \pa_{\rho} \eta \pa^{\rho} \nu - U(\vp ,\eta )
\lab{lagrange}
\er
where $h$ is the Coxeter number of $\lie$ and the potential being
\br
U(\vp ,\eta ) = \sum_{a=1}^r {4 q^a \o \a^2_a} e^{\qb (K_{ab}  \vp^b + \eta
)} + {4 q^0\o \psi^2} e^{\qb (-K_{\psi b}  \vp^b + \eta )}
\lab{potential}
\er
Let us introduce, for convenience the vector
\be
\vp \equiv \sum_{a=1}^r {2 \a_a \o \a_a^2} \vp^a
\lab{vectorphi}
\ee
The potential \rf{potential} can then be written as
\be
U(\vp , \eta ) = \sum_{j=0}^r {4 q^j\o \a_j^2} e^{\qb \left( \a_j.\vp +
\eta \right)}
\lab{potential2}
\ee
where we have denoted $\a_0 = - \psi$ as the extra simple root of the affine
Kac-Moody algebra ${\hat \lie}$.

The potential \rf{potential2} is invariant under the transformation
\be
\vp \rightarrow \vp + {2 \pi i \o \qb} \mu^{v} \, \, \, ; \, \, \,
\eta \rightarrow \eta + {2 \pi i \o \qb} n
\lab{degvacua}
\ee
where $\mu^{v}$ is a coweight of $\lie$, i.e. $\mu^v = \sum_{a=1}^r
m_a {2 \lambda_a \o \a_a^2}$ where $m_a$ and $n$ are integers. Such
transformation is complex and therefore in the regime where the coupling
constant $\qb$ is pure imaginary the real vacuum solutions are degenerate. This
generalizes to any simple Lie algebra the degenerate vacua of the sine Gordon
model where the minima of the potential are identified with (co-) weight
lattice of $SU(2)$.  Such degenerate vacua are responsible for the appearance
of topological soliton solutions in the AT and CAT models as we describe
below.

Notice that the potential \rf{potential2} and the CAT model equations of motion
\rf{todaone}-\rf{todathree} are invariant under the transformation
\be
\vp \rightarrow \vp + \zeta {\hat t} \, \, \, \, ; \, \, \, \,
\eta \rightarrow \eta - \zeta \, \, \, \, ; \, \, \, \,
\nu \rightarrow \nu + b \zeta
\ee
where $b$ is an arbitrary constant and $\zeta$ is an harmonic function, i.e.,
$\pa_{+}\pa_{-}\zeta = 0$, and
\be
{\hat t} \equiv \sum_{j=0}^r {2 {\hat \lambda}_j\o \a_j^2}
\ee
where ${\hat \lambda}_j$ are the fundamental weights of the Kac-Moody
algebra ${\hat \lie}$, i.e. ${\hat \lambda}_0 = (0, {\psi^2\o 2}
l_0^{\psi},0)$, ${\hat \lambda}_a = (\lambda_a,{\psi^2\o 2}l_a^{\psi} ,0)$,
$a=1,2,...,r$, with $\lambda_a$ being the fundamental weights of the simple
Lie algebra $\lie$ and the scalar product defined as ${\hat a}.{\hat b} \equiv
a.b + a_C b_D + a_D b_C$ for ${\hat a} = (a, a_C, a_D)$ and ${\hat b} =
(b, b_C, b_D)$. When evaluating the scalar product we consider the simple
roots $\a_j$ as $r+2$-component vectors, i.e. $\a_a \equiv (\a_a, 0, 0)$,
$a=1,2,...,r$ and $\a_0 = (-\psi , 0, 1)$.

The canonical energy-momentum tensor corresponding
to the CAT model Lagrangian \rf{lagrange} is given by
\br
\Theta_{\rho \sigma} &=& {\qb^2 \o 2} \sum_{a,b=1}^r {2 \o \a^2_a} K_{ab}
\left\lb \pa_{\rho} \vp^a \pa_{\sigma} \vp^b-
\h g_{\rho \sigma} \pa_{\mu} \vp^a \pa^{\mu} \vp^b \right\rb \nonumber\\
 &+& {\qb^2 \o 2} \sum_{a=1}^r {2 \o \a^2_a} \left\lb \pa_{\rho} \vp^a
\pa_{\sigma} \eta + \pa_{\sigma} \vp^a \pa_{\rho} \eta - g_{\rho \sigma}
\pa_{\mu} \vp^a \pa^{\mu} \eta \right\rb \nonumber\\
 &+& \qb^2 {h \o 2} \left\lb \pa_{\rho} \eta \pa_{\sigma} \nu +
\pa_{\sigma} \eta \pa_{\rho} \nu - g_{\rho \sigma} \pa_{\mu} \eta \pa^{\mu} \nu
\right\rb \nonumber\\
&+& g_{\rho \sigma}\left\lb \sum_{a=1}^r {4 q^a \o \a^2_a} e^{\qb (K_{ab} \vp^b
+\eta )} + {4 q^0\o \psi^2} e^{\qb (-K_{ab} \vp^b +\eta )} \right\rb
\lab{emtensor}
\er
where $\rho$, $\sigma =0,1$ are space-time indices ($\partial_0 \equiv
\partial_t$, $\partial_1 \equiv
\partial_x$) and $g_{00}=1$, $g_{11}=-1$, $g_{01}=g_{10}=0$.

One finds that this E-M tensor has a non-zero trace given by
\be
\Theta_{\rho}^{\rho} = 2 U(\vp ,\eta )
\lab{emtrace}
\ee
Since the CAT model  is conformally invariant its E-M tensor can be made
traceless. The usual procedure is to add a term $W_{\rho \sigma} =
\(\pa_{\rho} \pa_{\sigma} - g_{\rho \sigma} \pa^2 \) f$ with an arbitrary
function $f$.  Clearly the addition of such a term does not violate the
conservation of $\Theta_{\rho \sigma} $.  The trace of the extra term is given
by $W_{\rho}^{\rho} = - \pa^2 f$ and this fixes the choice of $f$ which renders
the modified E-M tensor traceless.  Accordingly one verifies that the modified
traceless E-M tensor of the CAT model is given by \ct{AFGZ}
\be
\Theta_{\rho \sigma}^{\rm CAT} = \Theta_{\rho \sigma} - \qb \(\pa_{\rho}
\pa_{\sigma} - g_{\rho \sigma} \pa^2 \) \( \sum_{a=1}^r {2\o \a^2_a} \vp^a + h
\nu\) \lab{cattensor}
\ee
The CAT model can be obtained via a Hamiltonian reduction procedure from the
two-loop WZNW model \ct{AFGZ}. The  modification of the E-M tensor described
above is equivalent to the one performed on the Sugawara tensor of the two-loop
WZNW model in order for reduction procedure to respect the conformal
invariance \ct{AFGZ}.

The Affine Toda model (AT) associated to a simple Lie algebra $\lie$ possesses
rank-$\lie$ fields $\vp^a$, $a=1,2,...$ rank-$\lie$; and it is not conformally
invariant. Its equations of motions correspond to eq. \rf{todaone} for  $\eta
=0$. As shown in \ct{CFGZ} the AT models constitute a ``gauge
fixed'' version of the CAT models. The $\eta$ field can be ``gauged away'' by a
conformal transformation \rf{ge} with $f^{\pr} = e^{\eta_{+}(x_{+})}$ and
$g^{\pr} = e^{\eta_{-}(x_{-})}$ where $\eta_{+}$ and $\eta_{-}$ are the
parameters of the solution of the free field $\eta$, i.e. $\eta (x_{+},x_{-})
=  \eta_{+}(x_{+}) +  \eta_{-}(x_{-})$. Therefore for every regular solution of
the $\eta$ field the CAT model defined on a space-time $(x_{+},x_{-})$
corresponds to the AT model with the extra field $\nu$ defined on a space-time
$({\tilde x}_{+},{\tilde x}_{-})$ where ${\tilde x}_{+} = \int^{x_{+}} dy_{+}
e^{\eta_{+}(y_{+})}$ and ${\tilde x}_{-} = \int^{x_{-}} dy_{-}
e^{\eta_{-}(y_{-})}$. Such connection allows one to calculate exactly the
perturbation of the $\eta$ field on the dynamics of the AT model \ct{CFGZ}.

For the particular solution $\eta = 0$ the CAT and AT are defined on the same
space-time and the dynamics of the fields $\vp^a$ are the same on both models.
In such case the E-M tensor of these theories are related by
\be
\Theta_{\rho \sigma}^{AT} = \Theta_{\rho \sigma}^{CAT}\mid_{\eta = 0} +
\qb \(\pa_{\rho}\pa_{\sigma} - g_{\rho \sigma} \pa^2 \) \( \sum_{a=1}^r {2\o
\a^2_a} \vp^a + h \nu\)
\lab{emcat/at}
\ee
This relation provides a very interesting insight on the role of the broken
conformal symmetry of the AT models. If one considers classical solutions of
soliton type which can be put at rest at some Lorentz frame, then the energy of
the solution can be interpreted as the rest mass of the soliton. Such mass
should be proportional to some mass scale of the theory. Due to the conformal
(scale) invariance  such mass scale does not exist in the CAT model and
therefore the soliton mass should vanish. Alternatively, this can be seen by
taking the matrix element of $\Theta^{CAT}_{\rho \sigma}$ between one soliton
states of momentum $p$ and $p^{\prime}$, obtaining \ct{GM}
\be
\langle p \mid \int dx \, \Theta^{CAT}_{\rho \sigma} \mid p^{\prime} \rangle =
A P_{\rho} P_{\sigma} + B \left( K_{\rho} K_{\sigma} - K^2 g_{\rho\sigma}
\right)
\lab{gellmann}
\ee
where $P\equiv {p+p^{\prime}\o 2}$ and $K\equiv p-p^{\prime}$. Relation
\rf{gellmann} follows from the symmetry of $\Theta^{CAT}_{\rho\sigma}$ and its
conservation $\partial^{\rho}\Theta^{CAT}_{\rho\sigma}=0$. If in the classical
limit, which is obtained by considering the tree diagrams of the vertex
\rf{gellmann}, $B$ has no poles at $K^2 = 0$, we shall have in the forward
direction
\be
\langle p \mid \int dx \, \Theta^{CAT}_{\rho \sigma} \mid p \rangle = A
p_{\rho} p_{\sigma}
\ee
and because $\Theta^{CAT}_{\rho \sigma}$ is traceless it follows we must have
either $p^2=0$ or $A=0$ for $p^2\neq 0$. Therefore $\Theta^{CAT}_{\rho \sigma}$
does not give any contribution for the case of a soliton of mass different from
zero ($p^2 \neq 0$).

However in the AT model the same is not true, and from \rf{emcat/at} one
observes that the contribution to the soliton mass in the AT models comes from
a total divergence contained in the second term of \rf{emcat/at}.  Indeed,
denoting by $M$ the soliton mass and $v$ its velocity (in units of light
velocity) one gets
\br
{M v \o {\sqrt{1 - v^2}}} &=& \int^{\infty}_{-\infty} dx \, \Theta^{AT}_{01}
\nonumber\\
&=& \qb \int^{\infty}_{-\infty} dx \pa_{x}\pa_{t} \( \sum_{a=1}^r {2\o
\a^2_a} \vp^a + h \nu\) \lab{divergence}\\
&=& \qb \pa_{t} \( \sum_{a=1}^r {2\o \a^2_a} \vp^a +
h\nu\)\mid_{-\infty}^{\infty} \nonumber
\er
This universal formula will be used to determine  the masses of the solitons we
will find below, using Hirota's method.
This result was reported independently by Olive \ct{olive4} and we learned
while typing this paper \ct{olive3} that his results was obtained using very
similar arguments to ours.

\section{Hirota's Method}
\setcounter{equation}{0}
We now construct solitons solutions for CAT and AT models associated to any
simple Lie algebra $\lie$ using the Hirota's method \ct{Hirota}. We introduce
the $\tau$-functions as \footnote{{}From \rf{fi} and \rf{ni} one sees that,
for $B=0$, the $\tau$-functions are primary fields of conformal weight
$(0,0)$. Notice this definition of the $\tau$-functions is related to that
of ref. \ct{CFGZ} by $\tau_j \rightarrow (\tau_j)^{l_j^{\psi}}$. Also, these
$\tau$-functions are related to the Leznov-Saveliev \ct{LS} solution of the CAT
model \ct{CFGZ}. Comparing \rf{phinu} with eqs. (14) and (15) of ref. \ct{CFGZ}
(where $\qb = 1$) one can write (on shell and for $\eta = 0$)
\br
\tau_a &=& e^{{2\o \psi^2}l_a^{\psi}\sigma + \va_a} \langle \l_{(a)}
\mid e^{K_{+}(x_{+})}M_{+}(x_{+}) M^{-1}_{-}(x_{-}) e^{-K_{-}(x_{-})} \mid
\l_{(a)} \rangle  \nonumber\\
\tau_0 &=& e^{{2\o \psi^2}\sigma}\langle \l_{(0)} \mid
e^{K_{+}(x_{+})}M_{+}(x_{+}) M^{-1}_{-}(x_{-}) e^{-K_{-}(x_{-})} \mid \l_{(0)}
\rangle \nonumber
\er}
\be
\varphi^a ={1\o \qb}\( - \ln {\tau_a \over \tau_0^{l_a^{\psi}}} + \va_a \)
\;\;\;\;\;\;\;\;
\nu ={1\o\qb} {2 \over {\psi}^2} \left( \sigma - \ln \tau_0 \right)
\lab{phinu}
\ee
with $a=1,2,...,$ rank-$\lie$, and where \ct{CFGZ}
\be
\va_a = \sum_{b=1}^{rank \lie} (RK)^{-1}_{ab} \ln \left( {q^0 l_b^{\psi} \over
q^b} \right)
\ee
where $R$ is the matrix with entries $R_{ab}=\delta_{ab} + n_b^{\psi}$,
with $n_b^{\psi}$ being the integers in the expansion $\psi =
n_a^{\psi} \a_a$, and so $ l^{\psi}_{a} =  {\a^{2}_{a}\o {\psi ^2}}n_a^{\psi}$,
and
\be
(R^{-1})_{ab} = \delta_{ab} - {n_b^{\psi} \over h}
\label{niceinv}
\ee
Substituting these definitions in \rf{todaone} and \rf{todathree} for $\eta =0$
one obtains that the resulting equations can be decoupled into
\br
\tg (\tau_j) &=& \beta l_j^{\psi} \left(  1 -\prod_{k=0}^{rank \lie}
\tau_k^{-K_{jk}} \right) \lab{tau1}\\
\pa_{+} \pa_{-} \sigma &=& \beta
\lab{tau2}
\er
where $K_{ij}$, $i,j=0,1,2,...,$ rank-$\lie$, is the extended Cartan Matrix of
the Affine Kac-Moody algebra $\hatlie$ associated to $\lie$ (obtained from the
ordinary Cartan matrix of $\lie$ by adding one extra row and column
corresponding to the simple root $\a_0 = - \psi$).
Furthermore
\be
\bigtriangleup (F) \equiv \pa_{+} \pa_{-} \ln F = {\pa_{+} \pa_{-} F \over F}
- {\pa_{+} F \pa_{-} F \over F^2}
\lab{triangle}
\ee
and
\be
\beta = {q^j \over l_j^{\psi}} e^{K_{jk}\va_k} \, \, \, \, \, \, \, \,
\mbox{for any $j=0,1,\ldots,r$}
\lab{defbeta}
\ee
is a constant independent of the index $j$ \ct{CFGZ}.
We have also set $\va_0 = 0$. In the calculation we have used the fact that
$l_j^{\psi}$, $j=0,1,2,...,$ rank-$\lie$, with $l_0^{\psi}=1$, constitute a
null vector of the extended Cartan matrix. Indeed,
$\sum_{j=0}^{r}K_{ij}l_j^{\psi}= {2 \a_i.\a_0\o
\a_0^2} +\sum_{a=1}^{r}{2 \a_i.\a_a\o \a_a^2}l_a^{\psi}=0$, since
$\a_0=-\psi$ and ${\psi\o \psi^2} = \sum_{a=1}^{r}l_a^{\psi}{\a_a\o
\a_a^2}$.

The solution to \rf{tau2} gives $\sigma$ as
\be
\sigma (x_{+}, x_{-}) = \beta x_{+} x_{-} + F(x_{+}) + G(x_{-})
\lab{defsigma}
\ee
with $F$ and $G$ being arbitrary functions.

Reference \ct{hollo} provided for the first time the Hirota's solution for
the Affine Toda models associated to $A_r \equiv SU(r+1)$.
The technical manipulations, required in order to write down the
consistent Hirota's equation for the AT model,
relied on the $\tau$-functions which exceeded by one the number of fields
physically associated to the model.
The origin of the extra $\tau$- function, namely $\tau_0$,
can be traced back, in view of the analysis in \ct{CFGZ}, to the  $\nu$ field
revealing intrinsic connection of the Hirota's equations \rf{tau1} to
the structure of the CAT model.

In the spirit of the Hirota's method we expand the $\tau$-functions in a formal
power series in a parameter $\epsilon$ as
\be
\tau_i = 1 + \epsilon \tau^{(1)}_i + ... + \epsilon^{N_i} \tau^{(N_i)}_i
\lab{Nsoliton}
\ee

We will be interested in solutions were the space-time dependence of the
$\tau$-functions is given by
\be
\t^{(n)}_i = \d^{(n)}_i \; e^{n \Gamma}
\lab{ansatz}
\ee
with
\be
\Gamma = \g_{+} x_{+} + \g_{-} x_{-} + \xi = \gamma ( x - v t ) + \xi
\lab{Gamma}
\ee
where $\d^{(n)}_i$ are constant vectors to be found from Hirota's equations,
and $\gamma_{+}={\gamma\o 2}(1+v)$, $\gamma_{-}={\gamma\o 2}(1-v)$ and $\xi$
are parameters of the solution.

The basic idea is to expand Hirota's eqs. \rf{tau1} in powers of $\epsilon$ and
solve them order by order. The method gives an exact solution if the series
truncates at some finite order in $\epsilon$. The actual value of
the parameter $\epsilon$ is irrelevant in the procedure and it can be set to
unity at the end of the calculation. Below we give a detailed procedure to
solve Hirota's equations. But before that we would like to discuss some general
properties of the method.

\begin{theor}
Let us suppose one has a solution of the Hirota's eqs. and let
us denote by $N_i$ the highest power of $\epsilon$ in the Hirota's expansion
of $\tau_i$, $i=0,1,2,\ldots,$ rank-$\lie$.
Then $N_i$ are the components of a null vector of the extended Cartan matrix:
\label{theor:power}
\end{theor}
\be
K_{ij}\, N_j = 0
\lab{hpower}
\ee
Since the zero eigenvalue of $K_{ij}$ is non degenerate and since $l_i^{\psi}$
is a null vector of $K$, it follows that $N_i = \kappa\, l_i^{\psi}$, where
$\kappa$ is some positive integer.

\proof
Since $K_{ii}=2$ and $K_{ij}\leq 0$ for $i\neq j$ one observes that by
multiplying both sides of \rf{tau1} by $\tau_j^2$ the powers of $\tau$'s will
all be positive. After multiplication the term of highest power in $\epsilon$
on the l.h.s. of \rf{tau1} will be $ \epsilon^{N_i K_{ii}}
\(\tau^{(N_i)}_i\)^2 \D \( \tau^{(N_i)}_i \) $  which is zero
because of the ansatz \rf{ansatz}.
Notice now that on the r.h.s there is only one \underbar{unique} term
contributing to the highest power of $\epsilon$ leading to
\be
\({\t^{(N_i)}_i}\)^{K_{ii}} \epsilon^{N_i K_{ii}} =
\prod_{k=0,k \ne i}^{rank \lie} \({\tau_k^{(N_k)}}\)^{-K_{ik}}
\epsilon^{-N_k K_{ik}}
\lab{proj}
\ee
Since the sum of the powers of $\epsilon$ on the r.h.s. of the above equation
should equal $N_i K_{ii}$ one obtains \rf{hpower} .$\Box$

Therefore a necessary condition for the expansion \rf{Nsoliton} to
truncate at some finite order (when ansatz \rf{ansatz} is used) is that the
matrix $K_{ij}$ appearing in the Hirota's eqs. \rf{tau1} should be singular.

Substituting the expansion \rf{Nsoliton} into Hirota's eqs. \rf{tau1} and
using \rf{ansatz} and \rf{Gamma} one obtains that the resulting equation in
order zero in $\epsilon$ is trivially satisfied whilst the first order equation
leads to
\be
L_{ij} \delta^{(1)}_j = \lambda \delta^{(1)}_i
\lab{nld}
\ee
where $L_{ij} = l_i^{\psi} K_{ij}$, and $\l = {\gamma_{+} \gamma_{-}\o
\beta}={\gamma^2 (1 - v^2)\o 4 \beta}$. Therefore the parameters
of the solution are restricted by the possible eigenvalues of the matrix
$L_{ij}$. As we shall show below, except for the algebras $SU(6p)$ and
$SP(3p)$ with $p$ a positive integer, the higher terms $\d^{(n)}$ $(n\geq 2)$,
are uniquely determined  by $\d^{(1)}$.
Therefore, if the eigenvalues are non degenerate
there can be at most rank-$\lie + 1$ one-soliton solutions. However if there
are degeneracies we can have many more solutions. Since a right null vector of
$K_{ij}$ is also a right null vector of $L_{ij}$ and vice versa, it follows
that there will always be a zero eigenvalue. This eigenvalue is not
degenerate and the corresponding soliton
solution is trivial in the sense that the $\vp$'s fields (but not
$\nu$) are constants.

\subsection{Soliton Masses}
{}From \rf{phinu} one gets
\be
\qb \left( \sum_{a=1}^r {2\o \a^2_a} \vp^a + h\nu \right) =
- \sum_{j=0}^r {2\o \a_j^2} \log \tau_j
+ {2\o \psi^2} h \sigma + \sum_{a=1}^r {2\o \a_a^2} \va_a
\ee
where we have used the fact that $\sum_{a=1}^r {2\o \a_a^2} l_a^{\psi} =
{2\o \psi^2}(h-1)$. Therefore from the relation \rf{divergence} one obtains
\be
{ M v \o \sqrt{1 -v^2}} = -\sum_{j=0}^r {2\o \a^2_j} { \dot{\t_j} \o \t_j}
\bv_{-\infty}^{\infty} + {2 h \o \psi^2} \dot{\sigma} \bv_{-\infty}^{\infty}
= -\sum_{j=0}^r {2\o \a^2_j} { \dot{\t_j} \o \t_j}
\bv_{-\infty}^{\infty}
\lab{masstwo}
\ee
where we have chosen the functions $F$ and $G$ in the definition
\rf{defsigma} in such way that the contribution to the mass from the
$\sigma$ field vanishes. Substituting \rf{Nsoliton} and \rf{ansatz} into
\rf{masstwo} one gets
\be
{ M v \o \sqrt{1 -v^2}} = \g v \sum_{i=0}^r {2\o \a^2_i}
{ \eps \d^{(1)}_i e^{\G} + \eps^2 \d^{(2)}_i 2 e^{\G}... +
\eps^{N_i} \d^{(N_i)}_i N_i e^{N_i \G} \o
1 + \eps \d^{(1)}_i e^{\G} + ... + \eps^{N_i} \d^{(N_i)}_i e^{N_i \G} }
\bv_{-\infty}^{\infty}
\lab{massthree}
\ee
Recalling that $\Gamma = \g (x - v t) + \xi$ we see that by taking $\g > 0$
the lower limit $x \rightarrow -\infty$
does not contribute and the limit $x \rightarrow \infty$ is finite.
For $\g < 0$ the converse happens. Therefore the mass is proportional to
$\mid \g \mid$.
{}From Theorem (\ref{theor:power}) \,we have $N_j = \kappa l_j^{\psi}$ where
$\kappa$ is a positive integer. Therefore
\be
{ M v \o \sqrt{1 -v^2}} = \mid \g \mid v \sum_{j=0}^r {2\o \a^2_j} N_j
=  \mid \g \mid v {2 h \kappa \o \psi^2}
\lab{massfour}
\ee
with $h$ being the Coxeter number of $\lie$.
Recalling that $ \l = \g^2 (1-v^2) /4\b$ we therefore arrive at the following
expression for the soliton masses in the AT model
\be
M =  {4 h \kappa  \o \psi^2} m  \sqrt{ \l}
\lab{massfinal}
\ee
where we have denoted $m \equiv \sqrt{\b}$. This mass formula is true for any
algebra and for any solution constructed using the ansatz \rf{ansatz}. Notice
that the masses of solitons associated with a given eigenvalue $\l$ are
quantized in units of ${4 h \o \psi^2} m \sqrt{ \l}$. In addition the soliton
masses are proportional to the masses of the fundamental particles of the AT
model.

\subsection{Soliton charges}
An important consequence of Theorem (\ref{theor:power}) is that the asymptotic
values of the $\vp$'s fields are always finite for the solutions constructed
through the ansatz \rf{ansatz}. {}From \rf{phinu}, \rf{Nsoliton} and
\rf{ansatz} we see that, for $\g > 0$, $\vp_a (-\infty ) = {\va_a\o \qb}$
and since $N_i = \kappa l_i^{\psi}$ with $l_0^{\psi}=1$ we see that the limit
of $\tau_a / \tau_0^{l_a^{\psi}}$ as $x\rightarrow \infty$ is finite and so
from \rf{phinu} $\vp_a (\infty )$ is also finite\footnote{We would like to
point out that for $x\rightarrow \pm \infty$, $\tau_a /\tau_0^{l_a^{\psi}}$
becomes a phase, i.e.  $\lim_{x\rightarrow \pm \infty} \mid \tau_a
/\tau_0^{l_a^{\psi}}\mid = 1$}.  For $\g < 0$ the limits are interchanged.
Since the masses of the solitons, as shown above are finite,
it follows that such asymptotic values of $\vp$'s are vacua of the
potential \rf{potential2}.
As discussed before, in the regime where $\qb$ is purely imaginary the real
minima of the potential are degenerate.
We then introduce the topological charge of the solitons as
($i\qt \equiv \qb$)
\be
Q \equiv {\qt \o 2 \pi} \int_{-\infty}^{\infty} dx \pa_x \vp
=  {\qt \o 2 \pi}\left( \vp (\infty ) - \vp (-\infty )\right)
\lab{charge}
\ee
The actual calculation of the charges presents some ambiguities which we now
discuss. {}From \rf{phinu}, \rf{vectorphi} and \rf{Nsoliton} we have
\be
\vp = \sum_{a=1}^r {2 \a_a \o \a_a^2} \vp^a =
-{1\o i\qt}\sum_{a=1}^r {2 \a_a \o \a_a^2} \log {\tau_a \o \tau_0^{l_a^{\psi}}}
+ \va =
-{1\o i\qt}\sum_{j=0}^r {2 \a_j \o \a_j^2} \log \tau_j + \va
\lab{vectortau}
\ee
where $\va = {1\o i\qt} \sum_{a=1}^r {2 \a_a \o \a_a^2} \va_a$.  The
$\tau$-functions are in general complex and therefore the logarithm function
may not be single valued. One could therefore use the prescription, as in
\ct{hollo}, that $\log z = \log \mid z \mid + i {\bar \theta}$ for $z = \mid z
\mid e^{i \theta}$, where $0 \leq {\bar \theta}< 2 \pi$ and ${\bar \theta} -
\theta = $ multiple of $2 \pi$ . However the implementation of  such
prescription requires also a criterion on the use of the identity $\log AB =
\log A + \log B$. The reason is that the result depends on the order in which
the bar prescrition and the identity are used. For instance, one could get
$\log AB = \log A + \log B = \log \mid A \mid + \log \mid B \mid + i{\overline
{\arg A}} + i{\overline {\arg B}}$ or $\log AB = \log \mid AB \mid +
i{\overline {(\arg A + \arg B)}}$. Due to such ambiguities the values of the
charges one obtains will differ by a sum of co-roots of $\lie$. Therefore it is
quite clear which coset in $\Lambda_W^v / \Lambda_R^v$  the charge lies (where
$\Lambda_W^v$ and $\Lambda_R^v$ denote the co-weight and co-root lattices of
$\lie$ respectively), but the determination of the possible values of the
charge, inside the coset, that a given soliton solution can have is quite
delicate. For this reason we will give below the charges of the solutions we
constructed, up to a sum of co-roots of $\lie$. Following \rf{charge} and
\rf{vectortau} we will use the formula (for $\g > 0$)
\br
Q = - {1 \o 2 \pi i} \lim_{x\rightarrow \infty} \sum_{a=1}^r
{2 \a_a \o \a_a^2} \log {\tau_a\o \tau_0^{l_a^{\psi}}}
= - {1 \o 2 \pi i} \sum_{a=1}^r {2 \a_a \o \a_a^2}
\log {\delta_a^{(N_a)}\o (\delta_0^{(N_0)})^{l_a^{\psi}}}
\lab{charge1}
\er
where $N_j$ is the highest power of $\epsilon$ in the Hirota's expansion of
$\tau_j$, $j=0,1,2...,r$. For $\gamma < 0$ the sign of the charge reverses.

\subsection{Recurrence Method}
\label{sec:recurrence}
We now describe a method to solve Hirota's equations recursively. We show how
to determine the higher $\delta^{(n)}$ ($n\geq 2$) from $\delta^{(1)}$.
We write Hirota's equation \rf{tau1} as:
\be
G_i = \b F_i
\lab{gibfi}
\ee
by introducing the auxiliary quantities:
\br
G_i & \equiv & \t^2_i \tg (\tau_i)   \lab{gi}\\
F_i & \equiv & l^{\psi}_i \left( \t^2_i -
\prod_{k=0,k \ne i}^{rank \lie} \tau_k^{-K_{ik}} \right) \lab{fi1}
\er
and possessing the following $\eps$ expansion:
\br
G_i &=& \epsilon G^{(1)}_i + \epsilon^{2} G^{(2)}_i + \ldots \nonu \\
F_i &=& \epsilon F^{(1)}_i + \epsilon^{2} F^{(2)}_i + \ldots \nonu
\er
Clearly \rf{gibfi} must hold for each order i.e.
\be
G_i^{(n)} = \b F_i^{(n)} \lab{gifi}
\ee
where both sides of \rf{gifi} can be rewritten as
\br
G_i^{(n)} &=& \pa_{+} \pa_{-} \t^{(n)}_i + {\cal A}_i^{(n)} \( \t^{(n-1)},
\ldots, \t^{(1)} \) \lab{gin} \\
F_i^{(n)} &=& L_{ij} \t^{(n)}_j + {\cal B}_i^{(n)} \( \t^{(n-1)}, \ldots,
\t^{(1)} \) \lab{fin}
\er
where $L_{ij} = l_i^{\psi} K_{ij}$ and ${\cal A}^{(n)}$ and ${\cal B}^{(n)}$
are sums of products of $\t$'s such that the sum of orders of $\t$'s for each
term is $n$.
{}From now on we choose the ansatz \rf{ansatz} for the $\tau$-functions.
Substituting \rf{gin} and \rf{fin} into \rf{gifi} we get
\be
\( L_{ij}  - n^2 \l \d_{ij} \) \, \d^{(n)}_j =  V_i^{(n-1)} \qquad\;
n=1, 2\ldots
\lab{divi}
\ee
where $\l \equiv \g_{+} \g_{-} / \b$ and where we grouped together the
lower order terms $\d^{(k)}$'s with $k<n$ into one term $V_i^{(n-1)} $
nonlinear in $\d$'s and defined by
\be
V_i^{(n-1)} \( \d^{(n-1)}, \ldots, \d^{(1)} \) =
{ {\cal A}^{(n)}_i  / \b  - {\cal B}^{(n)}_i \o / \exp {n \Gamma}} =
\l a_i^{(n-1)} - b_i^{(n-1)}
\lab{defvi}
\ee
For convenience we have introduced above new quantities $b_i^{(n-1)} =
{\cal B}^{(n)}_i /\exp {n\Gamma} $ and
$a_i^{(n-1)} = {\cal A}^{(n)}_i /\( \g_{+} \g_{-} \exp {n \Gamma}\) $.
Notice that $V_i^{(n-1=0)}=0$ due to:
\be
a_i^{(0)} =  b_i^{(0)} = 0    \qquad;\qquad i=0,\ldots , r
\lab{ab1}
\ee
Hence the expansion of the $\tau$ function must always start at the
first order with $\d_i^{(1)} $ being an eigenvector
of $L_{ij}$ matrix as we explained in \rf{nld}.
This forces the parameter $\l$ of Hirota's equation to be one of the
eigenvalues $\l^{\lb k \rb}$ of the $L_{ij}$ matrix.
Since $V_i^{(n-1)} $ depends on $\d^{(k)}$ with $k <n$ as seen from its
definition \rf{defvi}, equation \rf{divi} is a recursive relation and can be
used to determine higher order $\d$'s.
If $n^2 \l$ is not equal to any eigenvalue of $L_{ij}$ then the recurrence
relation becomes
\be
\d^{(n)}_i = {S^{(n)}}^{-1}_{ij} V^{(n-1)}_j \qquad; \quad
S_{ij}^{(n)} \equiv L_{ij} - n^2 \l \d_{ij}
\lab{dfromv}
\ee
If $n^2 \l$ equals one of the eigenvalues of $L_{ij}$ we have to be more
careful. In fact, such kind of ``degeneracy'' opens the way for constructing
new solutions of Hirota's equation. Expanding both $V^{(n-1)}_j $ and
$\d^{(n)}_i$
\be
\d_i^{(n)} = \sum_k d_{\lb k\rb}^{(n)} v^{\lb k\rb}_i \quad ; \quad
V_i^{(n-1)} = \sum_k c_{\lb k\rb}^{(n-1)} v^{\lb k\rb}_i
\lab{expand}
\ee
in terms of the eigenvectors $v^{\lb k \rb}_j $ of the $L_{ij}$ matrix:
\be
L_{ij} v^{\lb k \rb}_j = \l^{\lb k \rb} v^{\lb k\rb}_i
\lab{eigenv}
\ee
and plugging this expansion back into eq. \rf{divi} results in
\be
\left( \l^{\lb k \rb} - n^2 \l \right) d_{\lb k\rb}^{(n)} =
c_{\lb k\rb}^{(n-1)}
\lab{perturb}
\ee
So, if $\l^{\lb k \rb} \neq n^2 \l$ then $d_{\lb k\rb}^{(n)}$ is determined
from \rf{perturb}. However if $\l^{\lb k \rb} = n^2 \l$, $d_{\lb k\rb}^{(n)}$
is undetermined and we can choose it arbitrarily. Notice however that for
consistency $c_{\lb k\rb}^{(n-1)}$ has to vanish in such case (in fact, in
all cases where the kernel is non trivial we found that
$c_{\lb k\rb}^{(n-1)}$ consistenly vanishes). Such non trivial kernel of
$L_{ij} - n^2 \l \d_{ij}$ introduces new parameters in the procedure and we
will show below that it leads to extra soliton solutions of the AT and CAT
models.

As we are going to show in the following sections, the eigenvalues of
the matrix $L_{ij}$ for the classical algebras can be written as
\be
\l^{\lb j \rb} = 4 c_{\lie} \sin^2 \left( {j \pi \o h}\right) \, \, \, \, ;
\, \, \, \, j=0,1,2,...,r_{\lie}
\ee
where $c_{\lie}=1$ and $r_{\lie}= $ rank-$\lie$ for $\lie = SU(r+1)$, and
$Sp(r)$, and $c_{\lie}=2$ for $\lie = SO(2r)$ and $SO(2r+1)$, whilst
$r_{\lie}= $ rank-$\lie - 1$ for  $\lie = SO(2r+1)$ and $r_{\lie}= $
rank-$\lie -2$ for $\lie = SO(2r)$. $SO(2r+1)$ has an eigenvalue $\l = 2$ and
$SO(2r)$ has two eigenvalues $\l = 2$.

Consequently the kernel of $L_{ij} - n^2 \l \d_{ij}$ for the classical
algebras is non trivial when there exists $j$, $j^{\prime}$ and $n$ such that
$ \sin^2 \left( {j \pi \o h}\right) = n^2
\sin^2 \left( {j^{\prime} \pi \o h}\right)$. The solutions for such equation
are very scarce and can be obtained from  the following more general theorem
\ct{gillet} which we here present without proof.
\begin{theor}
The only solutions of the equation
\be
\sin \left( a \pi \right) = c \sin \left( b \pi \right)
\ee
with $a$, $b$ and $c$ non-zero rational numbers are when either $c=\pm 1$ or
$\{ \sin \left( a \pi \right) ,\sin \left( b \pi \right) \} \subset \{ 1, -1,
{1\o 2}, -{1\o 2} \}$.
\label{theor:gillet}
\end{theor}

Therefore the kernel is non trivial for $SU(6p)$ and $Sp(3p)$ only, with $p$ a
positive integer, (where in both cases $h = 6p$) with $j=3p$,
$j^{\prime}= p$ and $n=2$. We will show later that in such cases we obtain
new soliton solutions.

For the exceptional algebras one can check that the kernel of
$L_{ij} - n^2 \l \d_{ij}$ is always trivial.

Summarizing, using the above method we can find the $\delta^{(n)}_i$
recursively using \rf{dfromv} and/or \rf{perturb}.
We expect to find the soliton solutions for the cases where this series
truncates.
That is, we have to have:
\be
V^{(n-1)}_i = 0  \qquad {\rm for}\quad n > N_i
\lab{truncate}
\ee
since from now on all $\d^{(n)}_i$ will be zero.

\subsection{Calculation of $V^{(n-1)}_i$}
\label{sec:vn}
The calculation of $V_i^{(n-1)}$ defined in \rf{defvi} can be easily performed
using the Leibniz rule ${d^n (AB) \o dx^n} = \sum_{l=0}^n {d^l A \o dx^l}
{d^{n-l} B\o dx^{n-l}}$ and the fact that for an arbitrary function $H$ of the
tau-functions we have $H^{(n)}= d^n H/d\eps^n \v_{\eps=0}/ n! $. Let us first
give a derivation of $a_i^{(n-1)}$ used to determine $V^{(n-1)}_i$ in
\rf{defvi}. {}From \rf{triangle} and \rf{gi} we have $G_i = \tau_i \pa_{+}
\pa_{-} \tau_i - \pa_{+} \tau_i \pa_{-} \tau_i$. Therefore
\be
G^{(n)}_i = \g_{+} \g_{-} \sum_{l=0}^n \( (n-l)^2
- l (n-l)\)  \d_i^{(l)} \d_i^{(n-l)}  e^{n \Gamma}
\lab{gni}
\ee
Dividing by $\g_{+} \g_{-} \exp \( n \Gamma  \)$
and subtracting the term $ n^2 \d^{(n)}_i$ corresponding to
$\pa_{+} \pa_{-} \t_i$ we obtain
\be
 a_i^{(n-1)} =  \sum_{l=1}^{n-1} \( n^2 - 3n l +2 l^2\) \d_i^{(l)} \d_i^{(n-l)}
\lab{lain}
\ee
The final form of $V^{(n-1)}_i$ depends through $b^{(n-1)}_i$ on the Cartan
matrix and varies therefore from algebra to algebra.
However there are few generic terms which always appear in $F_i^{(n)}$.
As seen from \rf{fi1} these terms are: $(\t_i^2)^{(n)}$,
$(\t_{k-1} \t_{k+1})^{(n)}$, $(\t_i \t_j^2)^{(n)}$ and $(\t_i \t_j \t_k
)^{(n)}$. Therefore we give here the results for generic quadratic and cubic
terms in $\t$-functions.  Applying the same procedure as for the $G_i$ function
we arrive at
\be
(\t_{i} \t_{j})^{(n)}/ e^{n \Gamma} =
\d_{i}^{(n)}+ \d_{j}^{(n)} + \sum_{l=1}^{n-1}
\d_{i}^{(l)} \d_{j}^{(n-l)}
\lab{tt}
\ee
and similarly
\br
(\t_{i} \t_{j} \t_{k})^{(n)}/ e^{n \Gamma} &=&
\d_{i}^{(n)} + \d_{j}^{(n)} + \d_{k}^{(n)}
+ \sum_{l=1}^{n-1} \d_{i}^{(l)} \d_{j}^{(n-l)}
+ \sum_{l=1}^{n-1} \d_{j}^{(l)} \d_{k}^{(n-l)}\nonumber \\
&+& \sum_{l=1}^{n-1} \d_{i}^{(l)} \d_{k}^{(n-l)}
+ \sum_{l=1}^{n-1}\sum_{m=1}^{l-1}\d_{i}^{(m)} \d_{j}^{(l-m)} \d_{k}^{(n-l)}
\lab{ttt}
\er
Based on these results we will complete the
calculation of $V^{(n-1)}_i$ below for the relevant algebras.

\section{ $A_r \sim SU( r+1)$}
\setcounter{equation}{0}

The extended Cartan Matrix for $SU(r+1)$ is given as
\be
K= \left( \begin{array}{rrrrrrr}
2  & -1 & 0 & 0 & \ldots & 0 & -1\\
-1 & 2 & -1 & 0 & \ldots & 0 & 0 \\
0  & -1 & 2 & -1 & \ldots & 0 & 0 \\
\vdots & \vdots & \vdots & \vdots & \vdots & \vdots & \vdots \\
0  & \ldots & 0 & -1 & 2 & -1& 0\\
-1 & 0 & 0&\cdots & 0 & -1 & 2 \end{array} \right)
\label{cartsu}
\ee
and Hirota's equation \rf{tau1} reads
\be
\tau_j^{2}\tg (\tau_{j}) =  \b \( \tau_j^{2} -
\tau_{j+1} \tau_{j-1}\right) \, \, \, \, \, \, \, \mbox{for
$j=0,1,2,\ldots,r$}
\lab{sunhiro}
\ee
where, due to the periodicity of the extended Dynkin diagram, it is understood
that $\tau_{j+r+1} = \tau_{j}$.  The first order Hirota solution is obtained
from the eigenvectors of $L_{ij} = l_{i}^{\psi} K_{ij}$ as in \rf{nld}, namely
\be
L_{ij}v_{j} = \lambda v_{i}
\lab{L}
\ee
Since $l_i^{\psi} = 1$ for all $i$'s, this yields the system of equations
\be
 v_{j-1} - ( 2 - \l ) v_j + v_{j+1} = 0 \, \, \, \;\; \mbox{for
$j=0,1,...,r$} \lab{eigenvsunb} \\
\ee
with
\be
v_{j+r+1} = v_j
\lab{v}
\ee
Eq. \rf{eigenvsunb} can be recognized as the recurrence relation for the
Chebyshev polynomials with $2-\l = 2x$ (see appendix A). Therefore any linear
combination of the Chebyshev polynomials of type I and II satisfy
\rf{eigenvsunb}. The relation \rf{v} is equivalent, in fact, to the secular
equation for $L$ and determines the eigenvalues. Using the trigonometric
representation of the Chebyshev polynomials given in appendix A we get
\be
\l_j \,= \, 4 \sin^2 \( {j \pi \o r+1 }\)  \, \, \, \, \, \, \, \mbox{for
$j=0, 1,2,\ldots,r$}
\lab{suneigenv}
\ee
By inspection  $ \l_j = \l_{r+1-j}$ which clearly indicates that all
eigenvalues, apart from $\l_0 = 0$ and $\l_{(r+1)/2}=4$ (for even $r+1$), are
degenerate.  The corresponding eigenvectors are
\br
v_k^{\lb 0\rb} &=& 1   \nonu \\
v_{\lb 1 \rb\, k}^{\lb \l_l\rb} &=& \exp\({2\pi i l k \o r+1}\) \quad;\quad
v_{\lb 2 \rb\, k}^{\lb \l_l\rb} =\exp\(-{2\pi i l k \o r+1}\) \, \, \,
\, \, \, \, \mbox{$l=1,2,\ldots,\lb {r\o 2}\rb$} \nonu \\
v_k^{\lb \l_{(r+1)/2}\rb} &=& (-1)^k \quad {\rm for}~r+1~{\rm even}
\lab{sunvectors}
\er
where $k =0, 1,2,\ldots,r$ and
$$
\left\lb {r\o 2}\right\rb\; =\; \cases{ (r-1)/2, & if $r$ is odd ;\cr
r/2, & otherwise.\cr}
$$
Let us now apply our Hirota perturbation method to $SU(r+1)$. {}From
\rf{defvi}, \rf{lain} and \rf{tt} we find a closed expression for
$V^{(n-1)}_i$ to be
\be
V^{(n-1)}_j = - \sum_{l=1}^n \( \( 1 -  \l \( n^2 - 3n l +2 l^2\) \) \d_j^{(l)}
\d_j^{(n-l)}   - \d_{j+1}^{(n-l)} \d_{j-1}^{(l)} \)
\lab{sunv}
\ee
According to \rf{nld}, $\d^{(1)}$ has to be an eigenvector of $L$. We start
with the eigenvalues $\l_0 = 0$ and $\l_{{{r+1}\o{2}}} = 4$ (for ${r+1}$ even)
which are not degenerate, and so we have $\d^{(1)}_j = 1$ and $\d^{(1)}_j =
(-1)^j$ respectively. Substituting these vectors into
expression \rf{sunv} for $n=2$ we see that $V_j^{(1)}=0$. Consequently from
\rf{dfromv} we have $\d^{(2)}_j =0$, which leads to $V_j^{(2)}=0$ and
$\d^{(3)}_j =0 $ and so on.  Hence the Hirota perturbation series truncates
at first order. For $\l_0 =0$ we then get $\tau_j = 1 + e^{\Gamma}$, which is a
trivial solution in the sense that the $\vp$'s fields are constants (see
\rf{phinu}). For $\l_{{{r+1}\o{2}}} = 4 = {\gamma_{+} \gamma_{-}\o \b}$ (for
${r+1}$ even) we get
\be
\tau_j = 1 + (-1)^j e^{\Gamma}
\ee
According to \rf{massfinal} the mass of such solution is $M={8(r+1)\o \psi^2}m$
($r+1$ even).

For the remaining set of (degenerate) eigenvalues we consider the following
general linear combination:
\be
\d^{(1)}_j = y_1 \exp\({2\pi i lj \o r+1}\) +y_2 \exp\(-{2\pi i lj \o r+1}\)
\, \, \, \, \, \, \, \mbox{$l=1,2,\ldots,\lb {r\o 2}\rb$}
\lab{sund1}
\ee
where $j=0,1,2,...,r$, $y_1$ and $y_2$ are arbitrary constants.  The
calculation based on \rf{sunv} gives:
\be
V^{(1)}_j = - 2y_1 y_2 \( 1 -  \cos \({4\pi l \o r+1}\) \)=
- y_1 y_2 \l_l ( 4 - \l_l)
\, \, \, \, \, \, \, \mbox{$l=1,2,\ldots,\lb {r\o 2}\rb$}
\lab{sunv1}
\ee
where we used expression \rf{suneigenv} for the eigenvalues $\l_l$.
Rewriting \rf{sunv1} as $V^{(1)} = y_1 y_2 \l_l ( 4 - \l_l) v^{\lb 0\rb}$
and expanding $ \d^{(2)} = d^{(2)}_k v^{\lb k\rb}$ in terms of the
eigenvectors from \rf{sunvectors} we get after substitution into
\rf{perturb}
\be
 \d^{(2)} = y_1 y_2 \( 1- {1\o 4} \l_l\) v^{\lb 0 \rb}
\lab{sund2}
\ee
Substituting this into \rf{sunv} we find $V^{(2)}=0$ and consequently
$\d^{(3)}=0$, which in turn leads to  $V^{(3)}=0$ and $\d^{(4)}=0$.
One can verify that $V^{(n)}=0$ with $n \geq 4$. Therefore $\d^{(n)}=0$
 for $n \geq 3$.

To summarize, the general solution of Hirota's equations for $A_r \sim SU(
r+1)$ is given by the tau-function
\be
\t_j = 1 + \(y_1 \exp\({2\pi i lj \o r+1}\) +
y_2 \exp\(-{2\pi i lj \o r+1}\) \) e^\G + y_1 y_2 \( 1- {1\o 4} \l_l\)
e^{2\G}
\lab{suntau}
\ee
with $\G$ given by \rf{Gamma}, ${\gamma_{+} \gamma_{-}\o \b}= \l_l = 4 \sin^2
\left( {l \pi \o r+1}\right)$ and $l=1,...,[{{{r}\o{2}}}]$.
The solutions given in \rf{suntau} contain those obtained by Hollowood when
either $y_1 = 0$ or $y_2 = 0$ where the corresponding masses are given by eqns.
\rf{massfinal} with $\kappa = 1$, i.e.
\be
M_l ={{8(r+1)}\o{\psi^2}}m \sin({{l\pi}\o{r+1}})
\ee
However when considering both $ y_1 $ and $ y_2   \neq 0$ , the
solution truncates only in the second order in $\epsilon$ yielding a
solution with mass twice as large.  We should point out that in
such case we have a ``two-soliton-like'' solution surviving in
the static limit.

If we consider the topological charge given in \rf{charge1} we find that when
$y_1 =0$  the corresponding charge is
\be
Q_{-l} = -{1 \o{2\pi}}\sum_{a=1}^{r} {2\a_{a} \o {\a_{a}^2}}\left({- 2\pi
la\o{r+1}}\right) =  \omega_{r+1-l} +  \b_{-}
\label{Q1}
\ee
for some $\b_{-} \in \Lambda_R$. Taking now $y_2 =0$ the charge is
\be
Q_{l} = -{1 \o{2\pi}}\sum_{a=1}^{r} {2\a_{a} \o {\a_{a}^2}}\left({2\pi
la\o{r+1}}\right) =  \omega_{l} +  \b_{+}
\label{Q2}
\ee
for some $\b_{+} \in \Lambda_R$ and $l=1,..., [ {{r}\o{2}}]$, $\omega_i \,\,
i=1,...r$ are the fundamental weights of $SU(r+1)$.

The case of solution\rf{suntau}  when $y_1 ,y_2 \neq 0$, under
such prescription correspond to  charges lying in the root
lattice $\Lambda_R$.

\subsection{Second type of degeneracy for $SU(6p)$}

As noted in  section \ref{sec:recurrence}, for the eigenvalue $\l_p =1$ of
$SU(6p)$ the higher order $\d^{(n)}$'s ($n\geq 2$) are not uniquely determined
by $\d^{(1)}$ since the kernel of  $L_{ij} - n^2 \l \d_{ij}$
is not trivial for $n=2$. We therefore can add to $\d^{(2)}$ a term
proportional to the eigenvector corresponding to $\l_{3p} =4$, introducing a
new parameter in the solution.
The first order contribution to the tau-function corresponding to
$\l_{p} =1$ is
\be
\d^{(1)}_j = y_1 \exp\({2\pi i pj\o 6p}\) +y_2 \exp\(-{2\pi i pj\o 6p}\)
= y_1 \exp\({i \pi j\o 3}\) +y_2 \exp\(-{i \pi j\o 3}\)
\lab{sund1galois}
\ee
where $y_1$ and $y_2$ are arbitrary constants reflecting
the ordinary degeneracy between eigenvalues $\l_p$ and
$\l_{6p-p}$. {}From \rf{sunv1} we get
\be
V^{(1)}_j = - 3 y_1 y_2
\lab{sunv1galois}
\ee
and it is easy to see that the linear combination
\be
\d^{(2)} = {3 y_1 y_2 \o 4 } v^{ \lb \l=0 \rb} + z v^{ \lb \l_{3p} \rb}
= {3 y_1 y_2 \o 4 } + z (-1)^j
\lab{sund2galois}
\ee
for arbitrary $z$ satisfies the second order recurrence relation:
$ ( L - 2^2 \l_p I) \d^{(2)} = V^{(1)}$ recalling that $2^2 \l_p = \l_{3p}$.
Plugging now \rf{sund2galois} into \rf{sunv} we obtain
\be
V^{(2)}_j  = (-1)^{j+1} z \( \d^{(1)}_j + \d^{(1)}_{j+1} + \d^{(1)}_{j-1} \)
= - 2 z \( y_1 u_{\lb 2 \rb\, j}^{ \lb \l_{2p} \rb}  +
+ y_2 u_{\lb 1\rb\, j}^{ \lb \l_{2p} \rb} \)
\lab{sunv2galois}
\ee
hence $V^{(2)}_j$ is a linear combination of two eigenvectors
corresponding to the degenerate eigenvalue $\l_{2p} =3$.
{}From recurrence relation $ ( L - 3^2 \l_p I) \d^{(3)} = V^{(2)}$ we now
find $\d^{(3)}_j = z (-1)^j \d^{(1)}_j /3$.
One more step of recurrence procedure yields
$V^{(3)}= -z y_1 y_2 (-1)^j$ and $\d^{(4)}_j = (-1)^j z y_1 y_2/12$.
At this point the recurrence procedure terminates and the final solution
is given by
\br
\t_j &=& 1 + \(y_1 \exp\({i \pi j \o 3}\) +
y_2 \exp\(-{i \pi j \o 3}\) \) e^\G + \( {3 \o 4}  y_1 y_2 + (-1)^j z \)
e^{2\G}\nonu \\
 &+& {z\o 3}  (-1)^j \( y_1 \exp\({i \pi j\o 3}\) +y_2
\exp\(-{i \pi j\o 3}\)\) e^{3 \G}
+(-1)^j {z y_1 y_2\o 12} e^{4 \G}
\lab{suntaugalois}
\er
which of course reproduces \rf{suntau} for $ z \to 0$.
The mass in such case is given by the mass formula \rf{massfinal} with $\kappa
= 4$, namely $M = {96mp \o{\psi^2}}$.  The topological charge in this case can
be verified to lie in the coset $\omega_{3p} + \Lambda_R$ of
$\Lambda_W /\Lambda _R$.

\section{ Sp(r) Case}
\setcounter{equation}{0}
The Cartan matrix is  given by
\be
K= \left( \begin{array}{rrrrrrr}
2 & -2 & 0 & 0 & \ldots & 0 & 0 \\
-1 & 2 & -1 & 0 & \ldots & 0 & 0 \\
0 & -1 & 2 & -1 & \ldots & 0 & 0 \\
\vdots & \vdots & \vdots & \vdots & \vdots & \vdots & \vdots \\
0 & 0 & \ldots & 0 & -1 & 2 & -1 \\
0 & 0 & 0 & \cdots & 0 & -2 & 2 \end{array} \right)
\ee
$l_i^{\psi} =1$ for $i=0,1,2...,r$, and yields the following  Hirota's
equations
\br
\tau_0^{2} \tg (\tau_0) &=& \beta \left( \tau_0^{2} -
\tau_1^{2}\right) \nonu\\
\tau_a^{2}\tg (\tau_{a}) &=& \beta \left( \tau_a^{2} -
\tau_{a+1} \tau_{a-1}\right) \, \, \, \, \, \, \, \mbox{for
$a=1,2,...,r-1$} \nonu\\
\tau_r^{2}\tg (\tau_r) &=& \beta \left(\tau_r^{2} -  \tau_{r-1}^{2}\right)
\lab{hirospr}
\er
The corresponding eigenvalue equation $ L_{ij} v_j = K_{ij} v_j = \l v_i$
reads in components as
\br
(2-\l ) v_0 - 2 v_1 &=& 0  \lab{eigenvspra} \\
- v_{a-1} + ( 2 - \l ) v_a - v_{a+1} &=& 0 \, \, \, \;\; \mbox{for
$a=1,2,...,r-1$} \lab{eigenvsprb} \\
-2 v_{r-1} + (2 - \l ) v_r &=& 0 \lab{eigenvsprc}
\er
{}From \rf{nld}, $\d^{(1)}$ has to be one of the eigenvalues of $L \equiv K$.
Using the results of section \ref{sec:vn} the function $V^{(2)}_i$ from
\rf{defvi} reads
\br
V^{(2)}_0 (\d^{(1)}) &=& (\d^{(1)}_0)^2 - (\d^{(1)}_1)^2 \lab{vispra} \\
V^{(2)}_a (\d^{(1)}) &=& (\d^{(1)}_a)^2 - \d^{(1)}_{a+1} \d^{(1)}_{a-1} \, \,
\, \mbox{for $a=1,2,...,r-1$} \lab{visprb}\\
V^{(2)}_r (\d^{(1)}) &=& (\d^{(1)}_r)^2 - (\d^{(1)}_{r-1})^2 \lab{visprc}
\er
Our soliton construction is based on two results:
\begin{enumerate}
\item The eigenvalues of the extended Cartan matrix are $ \l_{j}
= 4 \sin^2 \( j\pi / 2 r \)$, $ j = 0, \ldots,r$
\item  $V_i^{(2)} (\d^{(1)}) = (1-x^2) (\d^{(1)}_0)^2 $, where $x = (2-\l)/2$
and $V_i^{(n)} = 0$ for $n \geq 3$.
\end{enumerate}

The proof of these results is based on the basic properties of Chebyshev
polynomials \ct{arfken,rivlin} listed in Appendix A.
Let us start on the proof by first introducing the variable $x = (2-\l)/2$
into \rf{eigenvspra}, \rf{eigenvsprb} and \rf{eigenvsprc} and
rewriting
\be
v_i = T_i (x) v_0 \qquad; \; i=0, \ldots,r    \lab{viti}
\ee
Now equations \rf{eigenvspra} and \rf{eigenvsprb} are trivially satisfied due
to the basic recurrence relation \rf{rec1}.
The remaining equation \rf{eigenvsprc} becomes
\be
x T_r (x) - T_{r-1} (x) = 0  \lab{lasteq}
\ee
This equation is not satisfied automatically by the Chebyshev's polynomials.
It is in fact equivalent to imposing the secular equation for the
$SP(r)$ Cartan matrix.
Using \rf{rec2} we see that \rf{lasteq} is equivalent to
\be
(1 - x^2 ) T_r^{\pr} (x) = 0
\lab{newlast}
\ee
So the $(r+1)$ solutions to \rf{lasteq} are
given by the extrema of $T_r$ from \rf{extrema}.
This corresponds to
\be
\l_j = 4 \sin^2 \( {j \pi \o 2r}\) \lab{lambdaspr}
\ee
which proves the first observation.
Next, substituting \rf{viti} into \rf{vispra} we get
\be
V^{(2)}_0 = (1-x^2) (\d^{(1)}_0)^2  \lab{lemma1}
\ee
Using \rf{rec3} we get for \rf{visprb}
\be
V^{(2)}_a = \( T_a^2 - T_{a+1} T_{a-1} \) (\d^{(1)}_0)^2 =
\h ( T_0 - T_2) (\d^{(1)}_0)^2 = (1-x^2) (\d^{(1)}_0)^2 \lab{lemma2}
\ee
Using \rf{lasteq} we get
\be
V^{(2)}_r = \( T_r^2 - T_{r-1}^2 \) (\d^{(1)}_0)^2 =
(1-x^2) T_r^2 (\d^{(1)}_0)^2 = (1-x^2) (\d{(1)}_0)^2     \lab{lemma3}
\ee
where we used the fact that eigenvalues correspond to extremas for $T_r$
(see \rf{tneta}).
Therefore \rf{lemma1} and \rf{lemma2} hold for any value of $x$, while
\rf{lemma3} is valid only when $x$ corresponds to an eigenvalue of the $Sp(r)$
Cartan Matrix. So the first part of point 2 is proved. We now prove the rest.
The eigenvector corresponding to $\l =0$ ($x=1$) is $v_i^{\l=0} = 1,\,
\forall i$ as can be seen from \rf{tnone} or direct inspection.
Therefore $V^{(2)}$ is proportional to $v^{\l=0} $.
{}From \rf{expand} and \rf{perturb} we see that $\d^{(2)}$ is also
proportional to $v^{\l=0} $
\be
\d^{(2)}= \({(1+x) \o 8} (\d^{(1)}_0)^2\) v^{\l=0}   \qquad x \ne 1
\lab{d2}
\ee
(especially all components are equal). Note, that $\d^{(2)}$ vanishes for
$\l=0$ ($x=1$), due to indeterminacy in \rf{perturb} we don't get this result
from \rf{d2}

{}From direct calculation we get
\br
V^{(3)}_0 &=& (2-\l) \d_0^{(1)} \d_0^{(2)} - 2 \d_1^{(1)}
\d_1^{(2)} \lab{vi3spra} \\
V^{(3)}_a &=& (2-\l) \d_a^{(1)}\d_a^{(2)}  - \d_{a-1}^{(1)}\d_{a+1}^{(2)}
-\d_{a+1}^{(1)}\d_{a-1}^{(2)}  \, \, \, \mbox{for
$a=1,2,...,r-1$} \lab{vi3sprb}\\
V^{(3)}_r &=& (2-\l) \d_r^{(1)} \d_r^{(2)} - 2 \d_{r-1}^{(1)}
\d_{r-1}^{(2)} \lab{vi3sprc}
\er
After factorizing the equal components of $\d^{(2)}$ we are left with the
eigenvalue equation for $\d^{(1)}$ and so $V^{(3)}$ vanishes.
{}From \rf{perturb} also $\d^{(3)}$ must vanish. Hence $V^{(4)}$ will
be the same as $V^{(2)}$ with $\d^{(1)}$ replaced by $\d^{(2)}$.
Again since all components of $\d^{(2)}$ are equal $V^{(4)}$ vanishes.
Since all $V^{(n)}$ are quadratic in $\d$'s and since the highest nonvanishing
$\d$ is $\d^{(2)}$ it follows that $V^{(n>4)}=0$.

{}From the above considerations we can write the general soliton
solution for $SP(r)$ as
\be
\t_i = 1 + T_i (x) \d^{(1)}_0 e^{\G} + {(1+x) \o 8} (\d^{(1)}_0)^2 T_i (1) e^{2
\G}
\qquad x \ne 1 \;\; i= 0, 1,\ldots,r
\lab{sprsoliton}
\ee
where $x= (2-\l)/2$ and $\l = \g_{+} \g_{-} /\b$.
Notices that for $x=-1$ the series breaks at the first order and this
corresponds to known solutions of $Sp(r)$ \ct{CFGZ}.
For $x=1$ we have
\be
\t_i = 1 + T_i (1) \d^{(1)}_0 e^{\G}
\lab{sprsoliton1}
\ee
and so all the $\vp$ fields are constant since all $\t$'s are equal.

Masses can be easily analyzed from the universal soliton mass formula
\rf{massfinal}.
Clearly, $\kappa =2$ for the new solutions in \rf{sprsoliton} and taking into
account the value of the Coxeter number for $Sp(r)$ ($h=2r$) we find
\be
M_j = {8 r \o \psi^2} m  \sqrt{\l_j} ={16 r \o \psi^2} m  \sin \({\pi j\o
2r}\) \qquad;\quad j=0,\ldots,r
\lab{massspr}
\ee
For solutions with $x=-1$ the mass would be half as large.

The topological charge defined in \rf{charge1} lies in the coroot lattice
$\Lambda_R^{v}$ for $x \neq -1$ since all $\tau$'s possess in common the same
coefficient of the highest power in $e^{\Gamma}$.  For $x=-1$ the corresponding
topological charges given from \rf{sprsoliton} is
\be
Q_m =  -{1 \o{2\pi}}\sum_{a=1}^{r} {2\a_{a} \o {\a _{a}^2}}\left({
\pi m}\right) = \omega_{1} + \hat \b_m
\ee
where $\omega_1$ is a fundamental weight of $Sp(r)$ leading to
the defining representation.  Again the solutions
\rf{sprsoliton} provide solitons with topological charges lying
in the coset $\Lambda^{v}_W/\Lambda^{v}_R$ where $\Lambda^{v}_W$
and $\Lambda^{v}_R$ denote the coweight and coroot lattices of
$Sp(r)$ respectively.

\section{$D_r= SO(2r)$}
\setcounter{equation}{0}
The case of $D_4 = SO(8)$ has a special interest and so we start by discussing
it separately.
\subsection{$D_4  = SO(8)$}
The matrix $L_{ij} = l_i^{\psi} K_{ij}$ is given by
\be
L = \left(
\begin{array}{rrrrr}
2 & 0 & -1 & 0 & 0 \\
0 & 2 & -1 & 0 & 0 \\
-2 & -2 & 4 & -2 & -2 \\
0 & 0 & -1 & 2 & 0 \\
0 & 0 & -1 & 0 & 2
\end{array} \right)
\ee
and the integers $l_{i}^{\psi}$ are $\left( 1,1,2,1,1\right)$. The eigenvalues
of the matrix $L$ are $(0,2,2,2,6)$ and the corresponding right eigenvectors
are
\br
v^{\lambda =0} = \left( \begin{array}{r}
1\\
1\\
2\\
1\\
1
\end{array} \right) \, ;
v^{\lambda =2}_{\lb 1\rb} = \left( \begin{array}{r}
1\\
-1\\
0\\
1\\
-1
\end{array} \right) \, ;
v^{\lambda =2}_{\lb 2\rb} = \left( \begin{array}{r}
1\\
1\\
0\\
-1\\
-1
\end{array} \right) \, ;
v^{\lambda =2}_{\lb 3\rb} = \left( \begin{array}{r}
1\\
-1\\
0\\
-1\\
1
\end{array} \right) \, ;
v^{\lambda =6} = \left( \begin{array}{r}
1\\
1\\
-4\\
1\\
1
\end{array} \right)
\er
The Hirota's equations read
\br
\tau_0^2 \tg ( \tau_0 ) &=& \beta \left( \tau_0^2 - \tau_2 \right) \nonumber \\
\tau_1^2 \tg ( \tau_1 ) &=& \beta \left( \tau_1^2 - \tau_2 \right) \nonumber\\
\tau_2^2 \tg ( \tau_2 ) &=& 2\beta \left( \tau_2^2 - \tau_0  \tau_1 \tau_3
\tau_4 \right) \nonumber\\
\tau_3^2 \tg ( \tau_3 ) &=& \beta \left( \tau_3^2 - \tau_2 \right) \nonumber\\
\tau_4^2 \tg ( \tau_4 ) &=& \beta \left( \tau_4^2 - \tau_2 \right)
\lab{hirotaso8}
\er
The solution corresponding to ${\gamma_{+} \gamma_{-}\o \beta} = \lambda = 6$
is given
\br
\delta^{(1)} &=& \left( 1,1,-4,1,1 \right) \nonumber\\
\delta^{(2)} &=& \left( 0,0,1,0,0 \right)
\lab{so8l=6}
\er
and all the remaining $\delta$'s vanish. Therefore according to Theorem
\ref{theor:power} we have $\kappa =  1$. So, from \rf{massfinal} we get the
mass of such soliton is $M = {2\o \psi^2}12 \sqrt{6} m$.

The eigenvalue $\l = 2$ has multiplicity $3$. We then apply Hirota's method by
taking $\delta^{(1)}$ to be a general linear combination of the three
eigevectors. The solution is then given by

\br
\delta^{(1)} = \left( \begin{array}{c}
{x_1} + {x_2} + {x_3}\\
-{x_1} + {x_2} - {x_3}\\
0\\
{x_1} - {x_2} - {x_3}\\
-{x_1} - {x_2} + {x_3}
\end{array} \right) \, \,\, ;
\delta^{(2)} = {1\over 3}
\left( \begin{array}{c}
{{x_1}\,{x_2}} +
   {{x_1}\,{x_3}} +
   {{x_2}\,{x_3}}\\
  {- {x_1}\,{x_2}  } +
   {{x_1}\,{x_3}} -
   {{x_2}\,{x_3}}\\
  3\left( {{{x_1}}^2} + {{{x_2}}^2} + {{{x_3}}^2}\right) \\
  {- {x_1}\,{x_2}  } -
   {{x_1}\,{x_3}} +
   {{x_2}\,{x_3}}\\
  {{x_1}\,{x_2}} -
   {{x_1}\,{x_3}} -
   {{x_2}\,{x_3}}
\end{array} \right) \nonu
\er
\br
\delta^{(3)} =
P_3
 \left( \begin{array}{c}
1\\
1\\
-16\\
1\\
1
\end{array} \right)  \, ;
\delta^{(4)} = P_4 \left( \begin{array}{c}
0\\
0\\
1\\
0\\
0
\end{array} \right)\, ;
\delta^{(5)} = \left( \begin{array}{c}
0\\
0\\
0\\
0\\
0
\end{array} \right) \, ;
\delta^{(6)} = P_3^2
\left( \begin{array}{c}
0\\
0\\
1\\
0\\
0
\end{array} \right)
\lab{so8l=2}
\er
and the higher $\delta$'s vanish and where $P_3 = {{x_1}\,{x_2}\,{x_3}}/{27}$
and $P_4 = \left( {{{{x_1}}^2}\,{{{x_2}}^2}} + {{{{x_1}}^2}\,{{{x_3}}^2}} +
{{{{x_2}}^2}\,{{{x_3}}^2}} \right) / 9$. The solution holds true for any value
(even complex) of the parameters $x_1$, $x_2$ and $x_3$ used in the linear
combination.

The properties of the solution seem to be  insensitive to the actual values of
the parameters $x$'s as long as they do not vanish. However when one or two of
them vanish the solution changes substantially. According to Theorem
\ref{theor:power} we have $\kappa = 1,2,3$ when two, one and none of the $x$'s
vanish respectively. Then from \rf{massfinal} the masses are given by
\br
\begin{array}{ll}
M_1 = {2\o \psi^2} 12 \sqrt{2}\, m \, \, \, &; \, \, \, \mbox{when two $x$'s
vanish}\nonumber \\
M_2 = {2\o \psi^2} 24 \sqrt{2}\, m \, \, \, &; \, \, \,
\mbox{when one of the  $x$'s vanishes} \nonumber \\
M_3 = {2\o \psi^2} 36 \sqrt{2}\, m \, \, \, &; \, \, \,
\mbox{when none of the  $x$'s vanishes}
\end{array}
\lab{so8mass}
\er
The topological charge  defined in \rf{charge1} yields for the solution
\rf{so8l=6}, $Q \in  \Lambda_R$, the root lattice of $SO(8)$.  For those
solutions given in \rf{so8l=2}, the topological charge depends
upon the parameters $x_1, x_2$ and $x_3$.  By explicit calculation it can be
shown that
\br
Q(x_1,0,0) &\in &\l_s + \Lambda_R  \nonumber \\
Q(0,x_2,0) &\in & \l_v + \Lambda_R   \nonumber \\
Q(0,0,x_3) &\in &  \bar \l_s + \Lambda_R  \nonumber \\
Q(x_1,x_2,x_3) &\in & \Lambda_R
\er
where $\l_v, \l_s$ and $\bar\l_s$ are the vector and the two spinor weights of
$SO(8)$ respectively.
\subsection{$D_r= SO(2r)$ for $r\geq 5$}
The $L$ matrix is here given by
\be
L= \left( \begin{array}{rrrrrrr}
2 & 0& -1 & 0 & \ldots & 0 & 0 \\
0 & 2& -1 & 0 & \ldots & 0 & 0 \\
-2 & -2 & 4& -2 & \ldots & 0 & 0 \\
0 & 0 &-2& 4& -2& 0 &\ldots \\
\vdots & \vdots & \vdots & \ddots & \vdots & \vdots & \vdots \\
0 &\ldots & -2& 4& -2& 0& 0 \\
0 &0 &\ldots & -2& 4& -2& -2\\
0 & 0 & \ldots & 0 & -1& 2 & 0\\
0 & 0 & 0 & \cdots & -1& 0& 2 \end{array} \right)
\ee
with the integers $l_i^{\psi}$ given by $(1,1,2,2,...,2,1,1)$. To analyze the
eigenvalue equation  $ L_{ij} v_j = \l v_i$ we follow the usual procedure of
first introducing variable $x = (4-\l)/4$ and expressing the components $v_i$
in terms of Chebyshev polynomials.  As usual the generic part of the eigenvalue
problem takes the form of the recurrence relations for Chebyshev polynomials
\rf{rec1}.  Trying to fit an ansatz in form of linear combination of Chebyshev
polynomials into all eigenvalue equations produces the following result (in
case $\l\ne 2$ or $x \ne \h$):
\br
v_1 &=& v_0\qquad v_a= 2\(U_{a-1}-U_{a-2}\)v_0 \quad;\quad a=2,3,\ldots,r-2
\nonu\\
v_r &=& v_{r-1} =\(U_{r-2}-U_{r-3}\)v_0  \lab{so2rv}
\er
with the consistency condition, playing the role of the secular equation,
\be
v_{r-2} = \( 4x -2 \) v_r \quad {\rm or} \quad (x-1) T^{\pr}_{r-1} (x) = 0
\lab{so2rsecular}
\ee
to be imposed on the solution \rf{so2rv}.
The solutions to \rf{so2rsecular} take the following form
\be
\l_k = 8 \sin^2 \( {k \pi \o 2(r-1) }\) \quad k=0,1,\dots,r-2 \quad (\l \ne 2)
\lab{so2rlambda}
\ee
The case $\l=2$ has to be treated separately. One finds that the eigenvalue
$\l =2$ is twofold degenerated with eigenvectors:
\be
v^{\lambda =2}_{\lb 1\rb} = \left( \begin{array}{r}
1\\
-1\\
0\\
\vdots\\
0
\end{array} \right) \, \,\, ; \quad
v^{\lambda =2}_{\lb 2\rb} = \left( \begin{array}{r}
0\\
\vdots\\
0\\
-1\\
1
\end{array} \right)
\lab{ltwovectors}
\ee
In case where $r-1$ is a multiple of $3$ ($r-1=3N$) degeneracy becomes
threefold and the extra eigenvector has components given in terms of three
components $v_0,v_1,v_{r-1}$ playing role of free parameters
\br
v_{3+3k} &=& v_{4+3k} = - v_{6+3k} = - v_{7+3k} = -(v_0 + v_1) \quad
k= 0,2,4,6,\ldots   \nonu\\
v_{2+3k} &=& 0 \qquad k= 0,1,,\ldots, N \nonu\\
v_r &=& - v_{r-1} + (-1)^{N+1} (v_0 + v_1) \lab{ltwoextra}
\er
The Hirota's equations read
\br
\tau_0^2 \tg ( \tau_0 ) &=& \beta \left( \tau_0^2 - \tau_2 \right) \nonumber \\
\tau_1^2 \tg ( \tau_1 ) &=& \beta \left( \tau_1^2 - \tau_2 \right) \nonumber\\
\tau_2^2 \tg ( \tau_2 ) &=& 2\beta \left( \tau_2^2 - \tau_0  \tau_1 \tau_3
 \right) \nonumber\\
\tau_a^2 \tg ( \tau_a ) &=& 2\beta \left( \tau_a^2 - \tau_{a-1}  \tau_{a+1}
 \right) \, \, \, \, a=3,4,...,r-3\nonumber\\
\tau_{r-2}^2 \tg ( \tau_{r-2} ) &=& 2\beta \left( \tau_{r-2}^2 - \tau_{r-3}
\tau_{r-1} \tau_r \right) \nonumber\\
\tau_{r-1}^2 \tg ( \tau_{r-1} ) &=& \beta \left( \tau_{r-1}^2 - \tau_{r-2}
 \right) \nonumber\\
\tau_r^2 \tg ( \tau_r ) &=& \beta \left( \tau_r^2 - \tau_{r-2} \right)
\lab{hirotaso2r}
\er
Using the results of section \ref{sec:vn} one obtains that the quantities
$V_i^{(n-1)}$ defined in \rf{defvi} are given by
\br
V_0^{(n-1)} &=& -\sum_{l=1}^{n-1} \left( 1 - \l (n^2-3nl+2l^2) \right)
\d^{(l)}_0 \d^{(n-l)}_0 \nonumber\\
V_1^{(n-1)} &=& -\sum_{l=1}^{n-1} \left( 1 - \l (n^2-3nl+2l^2) \right)
\d^{(l)}_1 \d^{(n-l)}_1 \nonumber\\
V_2^{(n-1)} &=& -\sum_{l=1}^{n-1}  \left( 2 - \l (n^2-3nl+2l^2) \right)
\d^{(l)}_2 \d^{(n-l)}_2 \nonumber \\
&+& 2 \sum_{l=1}^{n-1}\left( \d^{(l)}_0 \d^{(n-l)}_1  +
 \d^{(l)}_1 \d^{(n-l)}_3 +
 \d^{(l)}_0 \d^{(n-l)}_3  +
 \sum_{m=1}^{l-1}\d^{(m)}_0 \d^{(l-m)}_1\d^{(n-l)}_3 \right)
\nonumber\\
V_a^{(n-1)} &=& -\sum_{l=1}^{n-1} \left( \left( 2 - \l (n^2-3nl+2l^2) \right)
\d^{(l)}_a \d^{(n-l)}_a - 2 \d^{(l)}_{a-1} \d^{(n-l)}_{a+1} \right)
\, \, \, ; \, \, a=3,4,...,r-3 \nonumber\\
V_{r-2}^{(n-1)} &=& -\sum_{l=1}^{n-1} \left( 2-\l (n^2-3nl+2l^2) \right)
\d^{(l)}_{r-2} \d^{(n-l)}_{r-2}\nonumber \\
&+& 2 \sum_{l=1}^{n-1}\left(
 \d^{(l)}_{r-3} \d^{(n-l)}_{r-1} +
 \d^{(l)}_{r-1} \d^{(n-l)}_r +
 \d^{(l)}_{r-3} \d^{(n-l)}_r  +
 \sum_{m=1}^{l-1}\d^{(m)}_{r-3} \d^{(l-m)}_{r-1}
\d^{(n-l)}_r \right) \nonumber\\
V_{r-1}^{(n-1)} &=& -\sum_{l=1}^{n-1} \left( 1 - \l (n^2-3nl+2l^2) \right)
\d^{(l)}_{r-1} \d^{(n-l)}_{r-1} \nonumber\\
V_{r}^{(n-1)} &=& -\sum_{l=1}^{n-1} \left( 1 - \l (n^2-3nl+2l^2) \right)
\d^{(l)}_{r} \d^{(n-l)}_{r}
\lab{viso2r}
\er
Applying the recurrence method described in section \ref{sec:recurrence} one
can now construct the solutions.

For the eigenvalues $\l_k = 8 \sin^2{{k \pi \o {2(r-1)}}} = {\gamma_{+}
\gamma_{-} \o \b} \neq 2$, $k=1,2,...,r-2$, we obtain the solution
\br
\tau_0 &=& 1 + e^{\Gamma} \nonumber\\
\tau_1 &=& 1 + e^{\Gamma} \nonumber\\
\tau_a &=& 1 + 2{\cos{(2a-1)\theta_k}\o \cos{\theta_k}}e^{\Gamma}
+ e^{2\Gamma} \, \, \, ; \, \, \, a=2,3,...,r-2 \nonumber\\
\tau_{r-1} &=& 1 + (-1)^k e^{\Gamma} \nonumber\\
\tau_r &=& 1 + (-1)^k e^{\Gamma}
\lab{so2rsolution}
\er
where $\theta_k = {k \pi \o 2(r-1)}$ and where we have used the fact that the
Chebyshev polynomials satisfy $U_{a-1}(x_k) - U_{a-2}(x_k)
={\cos{(2a-1)\theta_k}\o \cos{\theta_k}}$ with $x_k = \cos{2\theta_k}$.
According to Theorem \ref{theor:power} we see that $\kappa = 1$ in this case
and therefore form \rf{massfinal} the masses are given by
\be
M_k = {2\o \psi^2} 8 (r-1) \sqrt{2} \, m \, \sin{{k \pi \o 2(r-1)}} \, \, \, ;
\, \, \, k=1,2,...,r-2
\lab{so2rmass}
\ee
These solutions correspond to topological charges lying in the coset $Q_k \in
\l_v + \Lambda_R$ for $k$ odd and $Q_k \in \Lambda_R$ for $k$ even,where
$\l_v$ denote the vector weight of $SO(2r)$.

For $\l = 2$ we have to consider the case when $r-1$ is a multiple of $3$
separately since as shown above the eigenvalue has multiplicity three. Let us
consider first the case $r-1 \neq $ multiple of $3$. The eigenvalue $\l = 2$
has multiplicity $2$ and the eigenvectors are given by \rf{ltwovectors}. So,
taking $\d^{(1)}$ to be a linear combination, with parameters $y_1$ and $y_2$,
of those eigenvectors and applying the procedure of section
\ref{sec:recurrence} we obtain the following solution
\br
\tau_0 &=& 1 + y_1 e^{\Gamma} + c(y_1, y_2) e^{2\Gamma} \nonumber\\
\tau_1 &=& 1 - y_1 e^{\Gamma} + c(y_1, y_2) e^{2\Gamma} \nonumber\\
\tau_a &=& 1 +d_a(y_1, y_2) e^{2\Gamma} + c(y_1, y_2)^2 e^{4\Gamma} \, \, ; \,
\, a=2,3,...,r-2 \nonumber\\
\tau_{r-1} &=& 1 - y_2 e^{\Gamma} + (-1)^{r-1} c(y_1, y_2) e^{2\Gamma}
\nonumber\\
\tau_r &=& 1 + y_2 e^{\Gamma} + (-1)^{r-1} c(y_1, y_2) e^{2\Gamma}
\lab{so2rsolutionl=2}
\er
where $\Gamma$ is defined in \rf{Gamma}, $\l = 2 = {\gamma_{+} \gamma_{-} \o
\b}$, $c(y_1, y_2) = (y_1^2 + (-1)^{r-1} y_2^2)/4(r-1)$ and
$d_a = (-1)^a \left( \left( 4(r-a) - 2 \right) y_1^2 + (-1)^r
(4a-2)y_2^2\right) /4(r-1)$, $a=2,3,...,r-2$.

There are two special solutions which correspond to the choices $y_2 = \pm i^r
y_1$ since in such case $c(y_1, y_2) = 0$. Then according to Theorem
\ref{theor:power} we have $\kappa = 1$ and from \rf{massfinal} the masses of
these two solutions are the same and equal to
\be
M_1 = {2\o \psi^2} 4(r-1) \sqrt{2}\, m \, \, \, ; \, \, \, \mbox{for $y_2 = \pm
i^r y_1$}
\ee
For $y_2 \neq \pm i^r y_1$ we have $\kappa =2 $ and therefore the mass is
\be
M_2 = {2\o \psi^2} 8(r-1) \sqrt{2}\, m \, \, \, ; \, \, \, \mbox{for $y_2 \neq
\pm i^r y_1$}
\ee
For $y_2 = i^ry_1$ we can show that tha topological charge for either $r=2k$ or
$r=2k+1$ lie in the cosets, $Q \in \l_s + \Lambda_R$ for even $k$ or $Q \in
\bar \l_s + \Lambda_R$ for  $k$ odd.  When $y_2 = -i^ry_1$ the situation is
reversed, i.e.  $Q \in \l_s + \Lambda_R$ for odd $k$ or  $Q \in \bar \l_s +
\Lambda_R$ for  $k$ even.

For the case where $r-1 =$ multiple of $3$, the eigenvalue $\l = 2$ has
multiplicity $3$. Analogously, we then take $\d^{(1)}$ to be a linear
combination of the three eigenvectors \rf{ltwovectors} and \rf{ltwoextra} and
apply the method described in section \ref{sec:recurrence}. The calculations
are a bit cumbersome and we give here the results for the simplest example
corresponding to $SO(14)$. The main properties of the solution for any
$SO(6p+2)$ ($p$ a positive integer) are the same for any $p$. Obviously as
described above, $SO(8)$ has also the eigenvalue $\l =2$ with multiplicity $3$.
However for $SO(8)$ the degeneracy of $\l =2$ is related to the symmetries
of the Dynkin diagram whilst for $SO(6p+2)$ ($p>1$) the degeneracy is
accidental. Indeed, the solutions are different.

\subsection{$SO(14)$ solutions for $\lambda = 2$}
The three eigenvectors for the eigenvalue $\lambda = 2$  are
\br
v^{\lambda = 2}_{\lbrack 1 \rbrack} &=& \left( 1,-1,0,0,0,0,0,0\right)
\nonumber\\
v^{\lambda = 2}_{\lbrack 2 \rbrack} &=& \left( 0,0,0,0,0,0,-1,1\right)
\nonumber\\
v^{\lambda = 2}_{\lbrack 3 \rbrack} &=& \left( 1,1,0,-2,-2,0,1,1\right)
\er
By taking $\delta^{(1)}$ as a linear combination of these eigenvectors and
applying the Hirota's procedure described  in section \ref{sec:recurrence} we
get the following solution
\br
\delta^{(1)} = \left( \begin{array}{c}
{y_1} + {y_3}\\
-{y_1} + {y_3}\\
0\\
-2\,{y_3}\\
-2\,{y_3}\\
0\\
-{y_2} + {y_3}\\
{y_2} + {y_3}
\end{array} \right) \, \, \, ;
\delta^{(2)} = \left( \begin{array}{c}
{{{{{y_1}}^2}}\over {24}} +
   {{{{{y_2}}^2}}\over {24}} +
   {{{y_1}\,{y_3}}\over 3}\\
  {{{{{y_1}}^2}}\over {24}} +
   {{{{{y_2}}^2}}\over {24}} -
   {{{y_1}\,{y_3}}\over 3}\\
  {{3\,{{{y_1}}^2}}\over 4} -
   {{{{{y_2}}^2}}\over 4} + {{{y_3}}^2}\\
  {{-7\,{{{y_1}}^2}}\over {12}} +
   {{5\,{{{y_2}}^2}}\over {12}} + {{{y_3}}^2}\\
  {{5\,{{{y_1}}^2}}\over {12}} -
   {{7\,{{{y_2}}^2}}\over {12}} + {{{y_3}}^2}\\
  {{-{{{y_1}}^2}}\over 4} +
   {{3\,{{{y_2}}^2}}\over 4} + {{{y_3}}^2}\\
  {{{{{y_1}}^2}}\over {24}} +
   {{{{{y_2}}^2}}\over {24}} -
   {{{y_2}\,{y_3}}\over 3}\\
  {{{{{y_1}}^2}}\over {24}} +
   {{{{{y_2}}^2}}\over {24}} +
   {{{y_2}\,{y_3}}\over 3}
\end{array} \right) \, ; \,
\delta^{(3)} = {y_3\over 27}\left( \begin{array}{c}
{{{{{y_1}}^2}} +
   {{{{y_2}}^2}}\over {8}}\\
  {{{{{y_1}}^2} } +
   {{{{y_2}}^2} }\over {8}}\\
  -2\left( \,{{{y_1}}^2}  +
   \,{{{y_2}}^2} \right) \\
  {{23\,{{{y_1}}^2} } -
   {13\,{{{y_2}}^2} }\over {2}}\\
  {{-13\,{{{y_1}}^2} } +
   {23\,{{{y_2}}^2} }\over {2}}\\
  -2\left( \,{{{y_1}}^2}  +
  \,{{{y_2}}^2} \right) \\
  {{{{{y_1}}^2} } +
   {{{{y_2}}^2} }\over {8}}\\
  {{{{{y_1}}^2} } +
   {{{{y_2}}^2} }\over {8}}
\end{array} \right) \nonumber
\er
\br
\delta^{(4)} = \left( \begin{array}{c}
0\\
0\\
P_4 +
   {{{{{y_1}}^2}\,{{{y_3}}^2}}\over {12}} -
   {{{{{y_2}}^2}\,{{{y_3}}^2}}\over {36}}\\
  P_4 -
   {{7\,{{{y_1}}^2}\,{{{y_3}}^2}}\over {108}} +
   {{5\,{{{y_2}}^2}\,{{{y_3}}^2}}\over {108}}\\
  P_4 +
   {{5\,{{{y_1}}^2}\,{{{y_3}}^2}}\over {108}} -
   {{7\,{{{y_2}}^2}\,{{{y_3}}^2}}\over {108}}\\
  P_4 -
   {{{{{y_1}}^2}\,{{{y_3}}^2}}\over {36}} +
   {{{{{y_2}}^2}\,{{{y_3}}^2}}\over {12}}\\
0\\
0
\end{array} \right)\, ; \,
\delta^{(5)} = P_5 \left( \begin{array}{c}
0\\
0\\
0\\
1\\
1\\
0\\
0\\
0
\end{array} \right) \, ; \,
\delta^{(6)} = P_6 \left( \begin{array}{c}
0\\
0\\
1\\
1\\
1\\
1\\
0\\
0
\end{array} \right)
\lab{so14solution}
\er

where $P_4 = \left( y_1^2 + y_2^2 \right)^2 /576$, $P_5 = - y_3 \left( y_1^2 +
y_2^2 \right)^2 / 2592$ and $P_6 = \left( y_3 \left( y_1^2 + y_2^2 \right) /
216 \right)^2$.

We should distinguish four type of solutions:
\begin{enumerate}
\item We get three solutions by taking $y_3 = 0$ and $y_2 =\pm i y_1$ or $y_3
\neq 0$ and $y_1 = y_2 = 0$. According to Theorem \ref{theor:power} we have
$\kappa = 1$ and then from \rf{massfinal} the masses are
\be
M_1 = {2\o \psi^2} 24 \sqrt{2} \, m
\ee
The topological charges for $y_2 = iy_1$ can be shown to lie in the coset $Q
\in \bar \l_s + \Lambda_R$ whilst for $y_2 = -iy_1$, $Q \in \l_s +
\Lambda_R$.  When $y_3 \neq 0$, and $y_1 =  y_2 = 0$, $Q \in
\Lambda_R$.
\item Taking $y_3 = 0$ and $y_2 \neq \pm i y_1$ we have $\kappa = 2$ and
then
\be
M_2 = {2\o \psi^2} 48 \sqrt{2} \, m
\ee
\item Taking $y_3 \neq 0$ and $y_2 =\pm i y_1$ again we have $\kappa = 2$
and then
\be
M_3 = {2\o \psi^2} 48 \sqrt{2} \, m
\ee
\item Finally when $y_3 \neq 0$ and $y_2 \neq \pm i y_1$we have $\kappa = 3$
and then
\be
M_4 = {2\o \psi^2} 72 \sqrt{2} \, m
\ee
\end{enumerate}
In the cases $2$, $3$ and $4$ above the topological charges lie in $\Lambda_R$.

\section{$B_r = SO(2r+1)$}
\setcounter{equation}{0}
\subsection{$SO(7)$}
The solitons solutions for the case $B_3 = SO(7)$ are very similar to those of
$SO(8)$. This is a consequence of the fact that the Dynkin diagram of $SO(7)$
can be obtained from that of $SO(8)$ by  a folding procedure. The Hirota's
equations can be obtained from \rf{hirotaso8} by making $\tau_4 \equiv \tau_3$
and ignoring the equation for $\tau_4$. The eigenvalues of the matrix $L$ for
$SO(7)$ are $(0,2,2,6)$.  The solution for $\l = 6$ for $SO(7)$ is obtained
from \rf{so8l=6} by deleting the last component of the vectors $\d^{(1)}$ and
$\d^{(2)}$. Since the integers $l_i^{\psi}$ are $(1,1,2,1)$ we get, according
to theorem \ref{theor:power}, that $\kappa = 1$ for this  solution. Then from
\rf{massfinal} the mass of the soliton is $M = {2\o \psi^2} 12 \sqrt{6}\, m$

The eigenvalue $\l = 2$ of $SO(7)$ has multiplicity $2$ and the solution is
constructed by taking $\d^{(1)}$ as a linear combination of the corresponding
eigenvectors and applying the procedures of section \ref{sec:recurrence}. The
solution one gets can be obtained from \rf{so8l=2} by making $x_3 \equiv x_1$
and by ignoring the last component of all $\d$'s (the solution will then depend
upon two parameters $x_1$ and $x_2$).  We have $\kappa = 1,2,3$ for $x_1 = 0$
and $x_2 \neq 0$, $x_1 \neq 0$ and $x_2 =0$ and $x_1\, ,\,  x_2 \neq 0$
respectively. The masses are then
\br
\begin{array}{ll}
M_1 = {2\o \psi^2}12 \sqrt{2}\, m \, \, \, &; \, \, \, \mbox{when $x_1 = 0$,
$x_2 \neq 0$ }\nonumber \\
M_2 = {2\o \psi^2}24 \sqrt{2}\, m \, \, \, &; \, \, \,
\mbox{when $x_1 \neq 0$, $x_2 = 0$} \nonumber \\
M_3 = {2\o \psi^2}36 \sqrt{2}\, m \, \, \, &; \, \, \,
\mbox{when $x_1\, ,\, x_2 \neq 0$}
\end{array}
\er
\subsection{$B_r = SO(2r+1)$ for $r\geq 4$}
The $L_{ij}=l^{\psi}_i K_{ij}$ matrix is here given by
\be
L= \left( \begin{array}{rrrrrrr}
2 & 0& -1 & 0 & \ldots & 0 & 0 \\
0 & 2& -1 & 0 & \ldots & 0 & 0 \\
-2 & -2 & 4& -2 & \ldots & 0 & 0 \\
0 & 0 &-2& 4& -2& 0 &\ldots \\
\vdots & \vdots & \vdots & \ddots & \vdots & \vdots & \vdots \\
0 &0 &\ldots & -2& 4& -2&0\\
0 & 0 & \ldots & 0 & -2& 4& -4\\
0 & 0 & 0 & \cdots & & 0-1& 2 \end{array} \right)
\ee
where the integers $l_i^{\psi}$ are $(1,1,2,2,...,2,1)$. To analyze the
eigenvalue equation  $ L_{ij} v_j = \l v_i$ we follow the usual procedure of
first introducing variable $x = (4-\l)/4$ and expressing the components $v_i$
in terms of Chebyshev polynomials.  As usual the generic part of the eigenvalue
problem takes the form of the recurrence relations for Chebyshev polynomials
\rf{rec1}.  Trying to fit an ansatz in form of linear combination of Chebyshev
polynomials into all eignvalue equations produces the following result (in case
$\l\ne 2$ or $x \ne \h$):
\br
v_1&=&v_0\nonumber \\
 v_a&=& 2\(U_{a-1}-U_{a-2}\)v_0 \quad;\quad a=2,3,\ldots,r-1
\nonu\\
v_r &=& \(U_{r-1}-U_{r-2}\)v_0  \lab{so2ronev}
\er
with the consistency condition, playing the role of the secular equation,
\be
v_{r-1} = \( 4x -2 \) v_r \quad {\rm or} \quad (x-1) T^{\pr}_{r} (x) = 0
\lab{so2ronesecular}
\ee
to be imposed on the solution \rf{so2ronev}.
The solutions to \rf{so2ronesecular} takes the following form
\be
\l_k = 8 \sin^2 \( {k \pi \o 2r}\) \quad k=0,1,\dots,r-1 \quad (\l \ne 2)
\lab{so2ronelambda}
\ee
The case $\l=2$ has to be treated separately. One finds that the eigenvalue
$\l =2$ has corresponding eigenvector:
\be
v^{\lambda =2}_{\lb 1\rb} = \left( \begin{array}{r}
1\\
-1\\
0\\
\vdots\\
\\
0
\end{array} \right) \lab{ltwovectorsodd}
\ee
In case where $r=3 (N+1)$ with some $N=1,2,\ldots$ we get degeneracy
with the extra eigenvector having components given in terms of two
components $v_0,v_1$ playing a role of free parameters
\br
v_{3+3k} &=& v_{4+3k} = - v_{6+3k} = - v_{7+3k} = -(v_0 + v_1) \quad
k= 0,2,4,6,\ldots  \nonu\\
v_{2+3k} &=& 0 \qquad k= 0,1,,\ldots, N \nonu\\
v_r &=& (-1)^{r/3} {(v_0 + v_1) \o 2} \lab{ltwoextraodd}
\er
The Hirota's equations read
\br
\tau_0^2 \tg ( \tau_0 ) &=& \beta \left( \tau_0^2 - \tau_2 \right) \nonumber \\
\tau_1^2 \tg ( \tau_1 ) &=& \beta \left( \tau_1^2 - \tau_2 \right) \nonumber\\
\tau_2^2 \tg ( \tau_2 ) &=& 2\beta \left( \tau_2^2 - \tau_0  \tau_1 \tau_3
 \right) \nonumber\\
\tau_a^2 \tg ( \tau_a ) &=& 2\beta \left( \tau_a^2 - \tau_{a-1}  \tau_{a+1}
 \right) \, \, \, \, a=3,4,...,r-2 \nonumber\\
\tau_{r-1}^2 \tg ( \tau_{r-1} ) &=& 2\beta \left( \tau_{r-1}^2 - \tau_{r-2}
 \tau_r^2 \right) \nonumber\\
\tau_r^2 \tg ( \tau_r ) &=& \beta \left( \tau_r^2 - \tau_{r-1} \right)
\lab{hirotaso2r+1}
\er
Using the results of section \ref{sec:vn} one obtains that the quantities
$V_i^{(n-1)}$ defined in \rf{defvi} are given by
\br
V_0^{(n-1)} &=& -\sum_{l=1}^{n-1} \left( 1 - \l (n^2-3nl+2l^2) \right)
\d^{(l)}_0 \d^{(n-l)}_0 \nonumber\\
V_1^{(n-1)} &=& -\sum_{l=1}^{n-1} \left( 1 - \l (n^2-3nl+2l^2) \right)
\d^{(l)}_1 \d^{(n-l)}_1 \nonumber\\
V_2^{(n-1)} &=& -\sum_{l=1}^{n-1}  \left( 2 - \l (n^2-3nl+2l^2) \right)
\d^{(l)}_2 \d^{(n-l)}_2 \nonumber \\
&+& 2 \sum_{l=1}^{n-1}\left( \d^{(l)}_0 \d^{(n-l)}_1  +
 \d^{(l)}_1 \d^{(n-l)}_3 +
 \d^{(l)}_0 \d^{(n-l)}_3  +
 \sum_{m=1}^{l-1}\d^{(m)}_0 \d^{(l-m)}_1\d^{(n-l)}_3 \right)
\nonumber\\
V_a^{(n-1)} &=& -\sum_{l=1}^{n-1} \left( \left( 2 - \l (n^2-3nl+2l^2) \right)
\d^{(l)}_a \d^{(n-l)}_a - 2 \d^{(l)}_{a-1} \d^{(n-l)}_{a+1} \right)
\, \, \, ; \, \, a=3,4,...,r-2 \nonumber\\
V_{r-1}^{(n-1)} &=& -\sum_{l=1}^{n-1} \left( 2-\l (n^2-3nl+2l^2) \right)
\d^{(l)}_{r-1} \d^{(n-l)}_{r-1}\nonumber \\
&+& 2 \sum_{l=1}^{n-1}\left(
 \d^{(l)}_{r} \d^{(n-l)}_{r} +
 2 \d^{(l)}_{r} \d^{(n-l)}_{r-2} +
 \sum_{m=1}^{l-1}\d^{(m)}_{r} \d^{(l-m)}_{r}
\d^{(n-l)}_{r-2} \right) \nonumber\\
V_{r}^{(n-1)} &=& -\sum_{l=1}^{n-1} \left( 1 - \l (n^2-3nl+2l^2) \right)
\d^{(l)}_{r} \d^{(n-l)}_{r}
\lab{viso2r+1}
\er
Applying the recurrence method described in section \ref{sec:recurrence} one
can now construct the solutions.

For the eigenvalues $\l_k = 8 \sin^2{{k \pi \o {2r}}} = {\gamma_{+}
\gamma_{-} \o \b} \neq 2$, $k=1,2,...,r-1$, we obtain the solution
\br
\tau_0 &=& 1 + e^{\Gamma} \nonumber\\
\tau_1 &=& 1 + e^{\Gamma} \nonumber\\
\tau_a &=& 1 + 2{\cos{(2a-1)\theta_k}\o \cos{\theta_k}}e^{\Gamma}
+ e^{2\Gamma} \, \, \, \, \, \, \, a=2,3,...,r-1\nonumber\\
\tau_r &=& 1 + (-1)^k e^{\Gamma}
\lab{so2r+1solution}
\er
where $\theta_k = {k \pi \o 2r}$ and where we have used the fact that the
Chebyshev polynomials satisfy $U_{a-1}(x_k) - U_{a-2}(x_k)
={\cos{(2a-1)\theta_k}\o \cos{\theta_k}}$ with $x_k = \cos{2\theta_k}$.
According to Theorem \ref{theor:power} we see that $\kappa = 1$ in this case
and therefore form \rf{massfinal} the masses are given by
\be
M_k = {2\o \psi^2} 8 r \sqrt{2} \, m \, \sin{{k \pi \o 2r}} \, \, \, ;
\, \, \, k=1,2,...,r-1
\lab{so2r+1mass}
\ee
and topological charges can be shown to lie in $Q \in \Lambda_R$.
Notice this solution can be obtained from \rf{so2rsolution} for $SO(2r+2)$ by a
folding procedure. Identifying $\tau_{r+1} \equiv \tau_r$ one observes that
\rf{so2rsolution} (for $r \rightarrow r+1$) reduces to \rf{so2r+1solution} and
the Hirota's equations \rf{hirotaso2r} reduces to \rf{hirotaso2r+1}.

Consider now the case where $r \neq $ multiple of $3$. The eigenvalue $\l =2$
is not degenerate in this case and taking $\d^{(1)}$ to be the eigenvector
\rf{ltwovectorsodd} and applying the procedure of section \ref{sec:recurrence}
one obtains the following solution
\br
\tau_0 &=& 1 +  e^{\Gamma} + {1\o 4r} e^{2\Gamma} \nonumber\\
\tau_1 &=& 1 -  e^{\Gamma} + {1\o 4r} e^{2\Gamma} \nonumber\\
\tau_a &=& 1 + (-1)^a \left( 1 - {4a-2 \o 4r} \right) e^{2\Gamma} +
{1 \o (4r)^2} e^{4\Gamma} \, \, ; \,\, a=2,3,...,r-1 \nonumber\\
\tau_r &=& 1 + {(-1)^r \o 4r} e^{2\Gamma}
\er
where $\Gamma$ is given by \rf{Gamma} with ${\gamma_{+} \gamma_{-} \o \b} = 2$.
According to theorem \ref{theor:power} we have $\kappa =2$ and therefore from
\rf{massfinal} the mass of such soliton is
\be
M = {2 \o \psi^2} 8 r \sqrt{2} \, m
\ee
and its topological charge lie in $Q \in \Lambda_R$.
Again, the above solution can be obtained from \rf{so2rsolutionl=2} for
$SO(2r+2)$ (i.e. $r \rightarrow r+1$) by making $y_1 = 1$ and $y_2 = 0$ and
identifying $\tau_{r+1} \equiv \tau_r$.

For the case where $r = $ multiple of $3$, the eigenvalue $\l =2$ has
multiplicity $2$. The solutions are obtained by taking $\d^{(1)}$ to be a
linear combination of the eigenvectors \rf{ltwovectorsodd} and
\rf{ltwoextraodd} and applying the procedures of section \ref{sec:recurrence}.
Like in the $SO(6p+2)$ case the calculations are cumbersome. One can in fact
obtain such solutions for $SO(6p+1)$ from those of $SO(6p+2)$ by a folding
procedure. For instance, the solution for $SO(13)$ can be obtained from
\rf{so14solution} for the case of $SO(14)$ by making $y_2 = 0$ and neglecting
the last component of all $\d^{(n)}$'s, $n=1,2,...,6$. We then obtain three
types of solutions: {\it a)} for $y_1 = 0$ and $y_3 \neq 0$ we have $\kappa =1$
and the mass is $M_1 = {2\o \psi^2} 24 \sqrt{2} \, m$; {\it b)} for $y_1 \neq
0$ and $y_3 = 0$ we have $\kappa =2$ and the mass is $M_2 = {2\o \psi^2} 48
\sqrt{2} \, m$ and finally {\it c)} for $y_1 \, , y_3 \neq 0$ we have $\kappa =
3$ and the mass is $M_3 = {2\o \psi^2} 72 \sqrt{2} \, m $.

\section{$G_2$}
\setcounter{equation}{0}
The matrix $L_{ij}= l_i^{\psi} K_{ij}$ in this case is  given by
\be
L = \left (\begin{array}{rrr}
2 & -1 & 0 \\
-2& 4 & - 6\\
0 & -1 & 2
\end{array}
\right)
\lab{ng2}
\ee
where the integers $l_i^{\psi}$ are $(1,2,1)$. The eigenvalues of $L$ are
$(0,2,6)$ and the corresponding eigenvectors are:
\be
v^{\l=0} = \left(\begin{array}{r}
1\\ 2\\ 1
\end{array}
\right)  \; ; \;
v^{\l=2} = \left(\begin{array}{r}
3\\ 0\\ - 1
\end{array}
\right)  \; ;\;
v^{\l=6} = \left(\begin{array}{r}
1\\ -4\\ 1
\end{array}
\right)
\lab{vg2}
\ee
The Hirota's equations for $G_2$ are
\br
\tau_0^{2} \tg (\tau_0) &=& \beta \left( \tau_0^{2} -
\tau_1 \right) \nonu\\
\tau_1^{2}\tg (\tau_{1}) &=& 2 \beta \left( \tau_1^{2} -
\tau_{0} \tau_{2}^3 \right) \nonu\\
\tau_2^{2}\tg (\tau_2) &=& \beta \left(\tau_2^{2} -  \tau_{1} \right)
\lab{hirotag2}
\er

For $\l=6$ the solution obtained, using ansatz \rf{ansatz} and the procedure of
section \ref{sec:recurrence}, is given by
\be
\d^{(1)} = \left(\begin{array}{r}
1\\ -4\\ 1
\end{array}
\right)  \; ; \;
\d^{(2)} = \left(\begin{array}{r}
0\\ 1\\ 0
\end{array}
\right)
\lab{g2lambda6}
\ee
We therefore have $\kappa = 1$ and according to \rf{massfinal} the mass is
\be
M_1 = {2 \o \psi^2} 12 \sqrt{6} \, m
\ee

For $\l =2$ the solution is given by
\br
\d^{(1)} &=& \left(\begin{array}{r}
3\\ 0 \\ -1
\end{array}
\right)  \; ; \;
\d^{(2)} = \left(\begin{array}{r}
1\\ 3\\ - 1/3
\end{array}
\right)  \; ; \;
\d^{(3)} = {1 \o 27} \left(\begin{array}{r}
1\\ -16\\  1
\end{array}
\right)  \nonumber\\
\d^{(4)} &=& \left(\begin{array}{r}
0\\ 1/3\\ 0
\end{array}
\right)  \; ; \;
\d^{(5)} = \left(\begin{array}{r}
0\\ 0\\ 0
\end{array}
\right)  \; ; \;
\d^{(6)} = \left({1 \o 27}\right)^2 \left(\begin{array}{r}
0\\ 1\\ 0
\end{array}
\right)
\lab{g2lambda2}
\er
For such solution we have $\kappa = 3$ and therefore from \rf{massfinal} the
mass is
\be
M_2 = {2 \o \psi^2} 36 \sqrt{2} \, m
\ee

Notice that these two solutions can be obtained from the $SO(8)$ solutions by a
folding procedure. Indeed, by identifying $\tau_1 \equiv \tau_3 \equiv \tau_4$
one observes that the Hirota's equations \rf{hirotaso8} become \rf{hirotag2}
(after the relabeling $\tau_1 \leftrightarrow \tau_2$). In addition, under
such identifications the solution \rf{so8l=6} becomes \rf{g2lambda6} and the
solution \rf{so8l=2} becomes \rf{g2lambda2} by seting $x_1 = x_2 = x_3 = 1$.
Both solutions yields topological charges lying in the co-root lattice of
$G_2$.

\section{$F_4$}

The $L$ matrix for $F_4$ is given by
\be
L = \left( \begin{array}{rrrrr}
2 & -1 & 0 & 0 & 0 \\
-2 & 4 & -2 & 0 & 0 \\
0 & -3 & 6 & -6 & 0 \\
0 & 0 & -2 & 4 & -2 \\
0 & 0 & 0 & -1 & 2
\end{array} \right)
\ee
The integers $l_{i}^{\psi}$ are $\left( 1,2,3,2,1 \right)$. The eigenvalues of
$L$ are $(0, 6+2\sqrt{3}, 6-2\sqrt{3}, 3+\sqrt{3}, 3-\sqrt{3})$ and none of
them are degenerate.

The Hirota's equations read
\br
\tau_0^2 \tg (\tau_0 ) &=& \beta \left(\tau_0^2 - \tau_1 \right) \nonumber \\
\tau_1^2 \tg (\tau_1 ) &=& 2\beta \left(\tau_1^2 - \tau_0 \tau_2 \right)
\nonumber \\
\tau_2^2 \tg (\tau_2 ) &=& 3\beta \left(\tau_2^2 - \tau_1 \tau_3^2 \right)
\nonumber \\
\tau_3^2 \tg (\tau_3 ) &=& 2\beta \left(\tau_3^2 - \tau_2 \tau_4 \right)
\nonumber \\
\tau_4^2 \tg (\tau_4 ) &=& \beta \left(\tau_4^2 - \tau_3 \right)
\er

We construct the solutions using the ansatz \rf{ansatz} and the procedure of
section \ref{sec:recurrence}. The results are listed below.

The solution for $\lambda = 6 + 2 \sqrt{3} = {\gamma_{+} \gamma_{-} \o\b}$ is
given by
\br
\delta^{(1)} &=& \left( 1, -2(2+\sqrt{3}), 3(3+2\sqrt{3}), -2(2+\sqrt{3}), 1
\right)
\nonumber\\
\delta^{(2)} &=& \left( 0,1,3\left( 3 + 2 \sqrt{3} \right) \, ,
  1 ,0\right) \nonumber\\
\delta^{(3)} &=& \left( 0,0,1,0,0\right)
\er
We have $\kappa = 1$ and therefore from \rf{massfinal} the mass of the soliton
is
\be
M_1 = {2 \o \psi^2} 24 \sqrt{6 + 2 \sqrt{3}} \, m
\ee

The solution for $\lambda = 6 - 2 \sqrt{3} = {\gamma_{+} \gamma_{-} \o\b}$ is
\br
\delta^{(1)} &=& \left( 1, -2(2-\sqrt{3}), 3(3-2\sqrt{3}), -2(2-\sqrt{3}), 1
\right)\nonumber \\
\delta^{(2)} &=&  \left( 0,1,3\left( 3 - 2\sqrt{3} \right) \, ,
  1,0\right) \nonumber\\
\delta^{(3)} &=&  \left( 0,0,1,0,0\right)
\er
Again we have $\kappa = 1$ and therefore the mass is
\be
M_2 = {2 \o \psi^2} 24 \sqrt{6 - 2 \sqrt{3}} \, m
\ee
The solution for $\lambda = 3 + \sqrt{3} = {\gamma_{+} \gamma_{-} \o\b}$ is
\br
\delta^{(1)} &=& \left( {{1 - {\sqrt{3}}}\over 2},1,0,-{1\over 2},
  {{-1 + {\sqrt{3}}}\over 4}\right) \nonumber\\
\delta^{(2)} &=& \left( a_1,2\,\left( 27 + 14\,{\sqrt{3}} \right) \,a_1,{9\over
{64}},
  2\,\left( 3 + 2\,{\sqrt{3}} \right) \,a_1,a_1\right) \nonumber\\
\delta^{(3)} &=& b_1 \left( 0,-2,8\,\left( 1 + {\sqrt{3}} \right) ,1,0\right)
\nonumber\\
\delta^{(4)} &=& {a_1^2}\left( 0,1,\left( 63 + 4\, . \,{3^{{5\over 2}}} \right)
, 1,0\right) \nonumber\\
\delta^{(5)} &=& \left( 0,0,0,0,0\right) \nonumber\\
\delta^{(6)} &=& {a_1^3}\left( 0,0,1,0,0\right)
\er
where
\br
a_1 &=&{{5 + {3^{{3\over 2}}}}\over
   {64\,\left( 71 + 41\,{\sqrt{3}} \right) }} \\
b_1 &=& {{-\left( 12417 + 7169\,{\sqrt{3}} \right) }\over
   {128\,\left( 172947 + 99851\,{\sqrt{3}} \right) }}
\er
In this case we have $\kappa = 2$ and therefore from \rf{massfinal} the mass of
the soliton is
\be
M_3 = {2 \o \psi^2} 48 \sqrt{3 +  \sqrt{3}} \, m
\ee

The solution for $\lambda = 3 - \sqrt{3} = {\gamma_{+} \gamma_{-} \o\b}$ is
\br
\delta^{(1)} &=& \left( {{1 + {\sqrt{3}}}\over 2},1,0,-{1\over 2},
  {{-\left( 1 + {\sqrt{3}} \right) }\over 4}\right) \nonumber\\
\delta^{(2)} &=& \left( a_2,2\,\left( 27 - 14\,{\sqrt{3}} \right) \,a_2,{9\over
{64}},
  2\,\left( 3 - 2\,{\sqrt{3}} \right) \,a_2,a_2\right) \nonumber\\
\delta^{(3)} &=& b_2 \left( 0,-2,8\,\left( 1 - {\sqrt{3}} \right) ,1,0\right)
\nonumber\\
\delta^{(4)} &=& {a_2^2}\left( 0,1,\left( 63 - 4\, . \,{3^{{5\over 2}}} \right)
, 1,0\right) \nonumber\\
\delta^{(5)} &=& \left( 0,0,0,0,0\right) \nonumber\\
\delta^{(6)} &=& {a_2^3}\left( 0,0,1,0,0\right)
\er
where
\br
a_2 &=& {{-5 + {3^{{3\over 2}}}}\over
   {64\,\left( -71 + 41\,{\sqrt{3}} \right) }} \\
b_2 &=& {{12417 - 7169\,{\sqrt{3}}}\over
   {128\,\left( -172947 + 99851\,{\sqrt{3}} \right) }}
\er
Again $\kappa = 2$ and the mass is
\be
M_4 = {2 \o \psi^2} 48 \sqrt{3 - \sqrt{3}} \, m
\ee
All solutions correspond to topological charges lying in the co-root lattice of
$F_4$.

\section{The $E_n$ series}
The one-soliton solutions for the exceptional algebras $E_6$, $E_7$ and $E_8$
were presented in \ct{mmc}. However not all static soliton solutions were
obtained there. In the case of $E_6$ the eigenvalues $\l_1 = 6-2\sqrt{3}$ and
$\l_2 = 6+2\sqrt{3}$ have multiplicity $2$. Therefore using our procedure of
taking $\d^{(1)}$ as a linear combination of the eigenvectors associated to the
degenerate eigenvalue, one obtains new solutions for $E_6$. We do not present
them here, but under a folding procedure they lead to the solitons of $F_4$,
constructed above, corresponding to the eigenvalues $6-2\sqrt{3}$ and
$6+2\sqrt{3}$. The masses of these new soliton solutions of $E_6$ are the same
as those of the corresponding solitons of $F_4$.
\section{Conclusions}

The Hirota tau-function method employed in this paper has proved to be
extremely powerful in providing explicit soliton solutions for the Toda
models.  In particular, when dealing with degenerate eigenvalues the method
prescribes how to consider a superposition of solutions for the non linear
problem.  This yields, for instance, new solutions not previously discussed in
the literature surviving in the static limit.

We have also given the massess of such solutions by using a mass formula
obtained by general arguments of conformal field theory.  For many of our
solutions we have discussed the topological charges and verified that they can
be classified according to the cosets $\L_w^v/\L_R^v$, the quotient of the
coweight and coroot lattices.

It is still unclear, however, how to use the representation theory of Lie
algebras in order to classify solutions and their topological
charges as conjectured by Hollowood in \ct{hollo}. A very interesting and
natural continuation of this work would be the study of the quantum theory of
these solitons. This could be approached from several directions like
bosonization, equivalence theorems and S-matrix \ct{andreas}.

\appendix
\section{Chebyshev Polynomials. Definitions and Identities}
\lab{appendixa}
\setcounter{equation}{0}
Here we list for completeness the basic properties of Chebyshev polynomials
\ct{arfken,rivlin}.
They play an instrumental role in solving the eigenvalue problem of the
$L_{ij}$ matrix for various Lie groups.

The defining relation for Chebyshev polynomials of both types,
$T_n (x)$ (type I) and $U_n (x) = T^{\pr}_{n+1}/(n+1)$ (type II) is given
in terms of recurrence relations:
\br
T_{n+1} (x) -2 x T_{n} (x) + T_{n-1} (x) &=& 0  \quad;\qquad n \geq 1 \nonu\\
U_{n+1} (x) -2 x U_{n} (x) + U_{n-1} (x) &=& 0
\lab{rec1}
\er
with $T_0 (x) = U_0 (x) = 1$ and $ T_{1} (x) = x$.
Another useful recurrence relations are:
\br
(1-x^2) T_{n}^{\pr} (x) &=&  -nx T_{n} (x) + n T_{n-1} (x)  \lab{rec2} \\
T_{m} (x) T_{n} (x) &= &\h \( T_{m+ n} (x) + T_{\v m- n\v} (x) \)
\qquad m,n \geq 0 \lab{rec3}
\er
The zeros $\xi_j$ of $T_n$: $ T_n ( \xi_j )=0 \;;\, j =1,2,\ldots,n$
are given by
\be
\xi_j = \cos { (2j-1) \o n} {\pi \o 2} \;\;\; ; \; \;\;  j =1,2,\ldots,n
\lab{zeros}
\ee
The extrema points $\eta_k$ at which $\v T_n (x) \v =1$ are given by
\be
\eta_k = \cos {k \pi \o n}  \;\;\; ; \;\;\; k=0, 1,2,\ldots,n
\lab{extrema}
\ee
At these points we also have
\be
T_n ( \eta_k) = (-1)^k \;\;\;\; ; \;\;  k=0, 1,2,\ldots,n
\lab{tneta}
\ee
It is clear that the points $\eta_1, \ldots, \eta_{n-1}$
are zeros of the derivatives $T_n^{\pr} (\eta_k)=0\; ; \; k=1,\ldots, n-1$
and therfore also zeros of $U_{n-1} (\eta_k)=0\; ; \; k=1,\ldots, n-1$.
One also has
\be
T_n (1) = 1 \qquad;\qquad T_n (-1) = (-1)^n
\lab{tnone}
\ee
The Chebyshev polynomials take very simple form in the trigonometric
representation:
\be
T_n (x) = \cos n \rho \quad;\quad U_n (x) = { \sin (n+1) \rho \o \sin \t}
\quad{\rm with} \quad x = \cos \rho
\lab{chebtrig}
\ee
One notices that for the special choice of arguments:
\be
\rho_j = {2 \pi j \o N+1} \qquad;\qquad j=0,1,2,\ldots,N
\lab{thetaj}
\ee
with some integer $N$ the Chebyshev polynomials satisfy the periodicity
relation
\be
T_{n+N+1} (\rho_j) = T_{n} (\rho_j) \quad;\quad U_{n+N+1} (\rho_j) = U_{n}
(\rho_j)
\lab{chebperiod}
\ee
Chebyshev polynomials become very useful in the study of the eigenvalue
equation $ L_{ij} v_j = \l v_i$ of the matrix $L_{ij} $ related to the Cartan
matrix.
In components the generic equation reads  as
\be
- v_{a-1} + ( 2 - \l ) v_a - v_{a+1} =0
\lab{cartangeneric}
\ee
for the parameter $a$ taking values between $2$ and $r-1 = {\rm rank}\, \lie
-1$.
Introducing variable $x = (2-\l)/2$
into \rf{cartangeneric} and rewriting $v_a$ as a linear combination of the
Chebyshev polynomials
\be
v_a = \( A T_a (x)+ B U_a (x) \) v_0
\lab{cheblinear}
\ee
we find that \rf{cartangeneric} is  trivially satisfied due
to \rf{rec1}.
The remaining equations of $ K_{ij} v_j = \l v_i$ (whose not appearing in
\rf{cartangeneric}) impose the secular equation on the $L$ matrix.
They have to be solved separately for each Lie group and solutions can be
found in terms of the zeros and extrema of the Chebyshev polynomials
as it was done in the text.

\lskip
{\bf Acknowledgements:}\\
One of us (LAF) acknowledges CNPq for financial support within CNPq/NSF
Cooperative Science Program and thanks Physics Department at the University of
Illinois at Chicago for hospitality.
We thank J.C. Campuzano, C. Halliwell and W. Poetz for providing computer
facilities.
We are very grateful to Prof. Henri
Gillet for showing us his proof of theorem (\ref{theor:gillet}).

\end{document}